\documentclass[traditabstract]{aa}
\usepackage{amssymb,amsmath,amsfonts}
\usepackage{graphicx,graphics}
\usepackage[usenames]{color}
\usepackage{ulem}

\newcommand{\beq}{\begin{equation}}
\newcommand{\eeq}{\end{equation}}
\newcommand{\bea}{\begin{eqnarray}}
\newcommand{\eea}{\end{eqnarray}}
\newcommand{\ks}{\kappa_{\rm S} }
\newcommand{\kt}{\kappa_{\rm T} }
\newcommand{\ras}{\mathrm{Ra_*} }

\newcommand{\rast}{Ra$_*$}

\newcommand{\nut}{{\mathrm{Nu_T}} }
\newcommand{\nutt}{Nu$_{\rm T}$ }
\newcommand{\nus}{{\mathrm{Nu_S}} }
\newcommand{\nust}{Nu$_{\rm S}$ }

\newcommand{\mr}{\mathrm}
\newcommand{\kse}{\kappa_\mr{s\, eff}}

\begin{document}
\title{Semiconvection: numerical simulations}
%\titlerunning{to be set}

\author{F.\ Zaussinger\inst{\ref{inst1},\ref{inst2},\ref{inst3}}, H.C.\ Spruit\inst{\ref{inst1}}}
\authorrunning{F.\ Zaussinger \& H.C.\ Spruit}

%\offprints{\\ H.\ Spruit, \email{henk@mpa-garching.mpg.de}}

\institute{
  Max-Planck-Institut f\"{u}r Astrophysik,
  Karl-Schwarzschild-Str.\ 1,
  D-85748 Garching, Germany \label{inst1}
\and 
Faculty of Mathematics, 
University of Vienna, 
Nordbergstra\ss e 15, 
A-1090 Vienna, Austria \label{inst2}
\and 
Dept. Aerodynamics and Fluid Mechanics, 
BTU Cottbus, 
Siemens-Halske-Ring 14, 
D-03046 Cottbus, Germany \label{inst3}
}
\date{\today}

\abstract{
A grid of numerical simulations of double-diffusive convection is presented for  the astrophysical case where viscosity (Prandtl number Pr) and solute diffusivity (Lewis number Le) are much smaller than the  thermal diffusivity. As in laboratory and geophysical cases convection takes place in a layered form. The proper translation between subsonic flows in a stellar interior and an incompressible (Boussinesq) fluid is given, and the validity of the Boussinesq approximation for the semiconvection problem is checked by comparison with fully compressible simulations. The predictions of a simplified theory of mixing in semiconvection given in a companion paper are tested against the numerical results, and used to extrapolate these to astrophysical conditions. The predicted effective He-diffusion coefficient is nearly independent of the double-diffusive layering thickness $d$.  For a fiducial main sequence model (15 $M_\odot$) the inferred mixing time scale is of the order $10^{10}$ yr. An estimate for the secular increase of $d$ during the semiconvective phase is given. It can potentially reach a significant fraction of a pressure scale height.
}

\maketitle

\section{Introduction}
In models of stellar structure, situations are found where the heavier products of nuclear burning provide stability to a zone which otherwise would be unstable to convective overturning. Such a zone, or part of it, would become convective if something managed to mix its composition (R.J.\ Tayler 1953). The question whether such a zone should be treated as if it were mixed or not has become known as the {semiconvection} problem. Answers to this question differ substantially. In practice, recipes are used containing a free parameter that allows the degree of mixing to be varied. Calculations in which such a parameter is adjusted to match observations are then called `with semiconvection'.  Commonly used prescriptions are those of Langer 1985 and Maeder 1997. 

The presence of a semiconvective zone has only a minor effect on the thermal structure of the star. The assumed amount of mixing of composition is critical, however, because the evolution of the star is sensitive to the precise distribution of products of nuclear burning with depth in the star. The main goal of a theory for semiconvection is thus a good determination of the rate of mixing. From the perspective of the stellar evolutionist, the theory would ideally provide a formula for the rates of mixing and  transport (the effective diffusivities), as functions of local thermodynamic state and composition, and their gradients.

In Spruit (1992, hereafter S92) such formulas were derived, adapting the known physics of  {double-diffusive convection}   (Turner 1979, 1985, Proctor 1981, Huppert \& Turner 1981, Schmitt 1994) to the case of a stellar interior (discussed first in this context by Spiegel 1969, 1972). The expression developed in S92 makes use of simplifications valid in the limiting case of very large Rayleigh number and very low solute diffusivity.  In a companion paper (hereafter S13), the analysis in S92 is extended to cover the more moderate conditions accessible with numerical simulations. In the following, mixing rates and their dependence on the intrinsic parameters governing the problem are measured with such simulations.  The predictions of S13 are tested against  these results, and then used to estimate the expected mixing rates in semiconvective zones of stars.

One of the predictions in S13 is the existence of a maximum density ratio (the ratio $R_\rho$ of stabilizing solute to destabilizing thermal stratification) for which a steady layered state is possible. In a slightly different guise, this limit also figures prominently in Proctor's (1981) analytic analysis. In this analysis, he proved that in the limit of vanishing solute diffusivity there exists a layered state at any Rayleigh number above the critical value for convection in the absence of a stabilizing solute gradient, provided the density ratio is below this critical value. The model in S13 does not predict  what happens for density ratios just above or below this maximum value. The numerical results presented in Sect. \ref{max} clarify how the system behaves in this case.

  The development of a linear initial gradient into the final state of overturning layers separated by diffusive steps is studied with a few examples. This transient process shows the `Kato-oscillations' expected from linear theory {(see Sect.\  5 for an example)}. It is not very relevant for astrophysical application, however, since the transition to the layered state happens on a time scale (a finite number {of} buoyancy oscillations) that is negligible compared with the time scales of interest, and is bound to depend on the details of the initial state. Instead, the focus here is on the transport of heat and solute in a double diffusive staircase with layers of given thickness, as a function of the intrinsic parameters of the problem.
 
 {In Sect.\  2 the known physics of double diffusive convection of the semiconvective type is reviewed, in general and from an astrophysical perspective. Sect. 3 describes the transport properties of a single double-diffusive layer in terms of the model of S13. The numerical methods are given in Sect.\  4. In Sect.\  5 the results of our parameter study are shown and compared with the predictions of this model. Application to the case of semiconvection in a $15\,M_\odot$ main sequence star is discussed in Sect.\  6.}

\section{Semiconvection and double-diffusive convection}
\label{semidd}
Situations where a fluid is stabilized by the density gradient due to a dissolved heavy constituent occur in nature. An example is convection {in arctic oceans (fresh meltwater cooled from above, stabilized by the salts dissolved in the sea water, see the review in Schmitt 1994}). Intensively studied are East-African rift lakes (lakes Kivu, Nyos and Mounon, cf.\ Schmid et al. 2010). These are heated from below by volcanic activity, which also is a source of dissolved gases (carbon dioxide and methane, hereafter the `solute'). Their density stratification is stabilized against convection by the stable gradient resulting from the weight of the carbon dioxide. Efforts to prevent catastrophic release of carbon dioxide (Lake Nyos, e.g.\ Sigvaldason 1989)   or to enable safe commercial exploitation   of methane (Lake Kivu, Nayar 2009) have led to extensive study of the fluid flows, heat flux and mixing rates in these natural double-diffusive systems.

The gradients in temperature and solute in the East-African lakes and the arctic are observed to be `stepped': consisting of a stack {(called `staircase')} of thin layers (decimeters to decameters). Inside a layer, overturning convection keeps the composition nearly uniform, with stable gradients in temperature and composition separating the layers. The physics involved is easily reproduced under controlled laboratory conditions (Turner \& Stommel 1964, Turner 1985).\footnote{Also on a coffee table. Latte macchiato in a tall glass often shows the effect nicely. After the coffee is added to the milk, a stably stratified gradient of milk/coffee mix develops (showing internal gravity waves in the form of a sloshing motion with a period of a few seconds). After about a minute, the initially smooth gradient starts dividing into thin (a few mm) layers, visible at low contrast. In the course of several minutes these merge into a smaller number of more clearly defined layers.} The layers are very long-lived:  of the order of months or more in the geophysical examples mentioned,  orders of magnitude  longer than the convective turnover times inside the layers.

In the stable gradients between the overturning layers the transport of the stabilizing solute takes place by diffusion instead of convection. This strongfly limits the effective transport of solute through the double-diffusive staircase. Residence times on the order of 1000 yrs are inferred for the solutes in lake Kivu, for example (Schmid et al. 2010). This is 8 orders of magnitude longer than the convective turnover times in these layers. The transport of heat is also strongly reduced; this is exploited for heat storage in solar ponds (cf. Lu and Swift, 2001). Similarly low fluxes have been measured in thermohaline staircases in the arctic and antarctic oceans (e.g.\ Padman \& Dillon 1987).

Theoretically, the observed layered nature of double-diffusive systems is well understood. Central to this understanding is the fact that linear stability analysis does not provide relevant clues to their behavior, because the double-diffusive case of thermal convection stabilized by a slowly-diffusing solute is {\it{subcritical}}. That is,  a nonlinear form of the instability, in the form of a  stable overturning flow, exists already {below} the critical temperature gradient for onset of linear instability. {The linear stability condition is thus not relevant for the behavior of the system (cf.\ Schladow et al.\ 1987)}.  

Linear instability predicts internal gravity modes to set in above some critical value of the temperature gradient, growing in amplitude by the effect of thermal diffusion: the so-called Kato oscillations (Kato 1966). Such oscillations (cf. movie at Fig.\ \ref{kato}) transport a negligible amount of heat or solute, compared with overturning motions of the same amplitude. For this reason alone, linear stability arguments cannot be used for estimates of the mixing rate in semiconvective zones. {More important, however, is the subcritical nature of double-diffusive convection. Proctor (1981) shows analytically that, in the limit of vanishing diffusivity of the solute, the layered form of convection exists whenever the Rayleigh number exceeds the critical value for ordinary convection, irrespective of the strength of the stabilizing component.  This assumption of vanishing solute diffusivity is eminently satisfied in the astrophysical case and holds reasonably well in geophysical and laboratory experiments.}

{In Stevenson 1979 it was assumed that the nonlinear development of the overstable oscillations would lead to saturation of the wave at a finite amplitude. This has been the rationale for some prescriptions used in astrophysics (e.g.\ Langer et al.\ 1985). The assumption of saturation at finite amplitude is appropriate for the more common supercritical forms of oscillatory instability, but not for the subcritical case, where a finite amplitude state exists for parameters where the system is still linearly stable. The assumption was also in conflict with observations: the laboratory and geophysical systems all showed the same characteristic form of convection in a system of overturning layers (e.g.\ Turner \& Stommel 1964), none that settled into finite amplitude oscillations. } 

In the literature on semiconvection it is often argued that the Prandtl number in astrophysics is much lower than in the geophysical and laboratory cases. The implied assumption that the understanding of double diffusive convection developed in these contexts can be set aside, is not necessary however. Proctor's analysis, for example, is largely independent of Prandtl number. It is valid as long as Pr is  not larger than of order unity, and viscosity is larger than the solute diffusivity. This is also satisfied in the astrophysical case.

{In physical terms, the} reason for the subcritical behavior can be understood with an energy consideration {(S92)}.   The amount of energy it takes to overturn a layer of thickness $d$ against a stable gradient scales as $d^2$ (as in a harmonic oscillator at the buoyancy frequency of the stratification). The expense in initial energy needed per unit mass to put the system into its finite-amplitude, layered state thus vanishes as $d$, down to some value where damping losses stabilize the system. A small initial perturbation, or an initial Kato oscillation, is sufficient to provide the energy for overturning into thin layers. Once established, this layered state is a stable form of convection. {In fact, reproduction of the oscillatory phase in the laboratory requires very careful setup of the initial stratification (Shirtcliffe 1967, see also the discussion in Huppert \& Turner 1981).} 

This agrees with the observation in laboratory experiments and geophysical systems like lake Kivu mentioned above, i.e. that the layering first sets in at a small thickness (cf. footnote above). In this context, the formation of layers found in the simulations by Rosenblum et al. 2011 is in line with expectation. Much less well-defined is the evolution on longer time scales, in particular the question how and on which time scales the layer thickness evolves (see Sect.\  \ref{merge} and S13).

\subsection{Double diffusive convection in stars}

Double diffusive convection in stars has traditionally been regarded as a piece of physics to be treated separately from the geophysical examples, since {the numerical} values of controlling parameters such as the Prandtl number are quite different. Apart from the difference in equation of state, however, the hydrodynamic equations are identical. Differences in physics that might be present between the two cases are in fact not apparent in the elementary recipes for semiconvection used in stellar evolution codes. The failure of these recipes when applied to the geophysical case is traditionally not considered an argument against their application in stars. {Such recipes include (a) to assume that no mixing takes place at all (the `Ledoux' recipe), another (b) to ignore the stabilizing effect of the solute gradient (`Schwarzschild' recipe, yielding a very high mixing rate) (c) some interpolation between these recipes, (d) to assume that the amount of mixing is such that the layer becomes marginally stable to overturning, and (e) the somewhat more physically motivated oscillation-based recipe of Langer et al.\ above.}  

Define the Prandtl number $\mathrm{Pr}=\nu/\kappa_{\rm T}$  and the  Lewis number $\mathrm {Le}=\kappa_{\rm S}/\kappa_{\rm T}$, where $\kappa_{\rm T}$ is the thermal diffusivity and  $\kappa_{\rm S}$ the diffusivity of the stabilizing `solute' ($\mathrm{He}$ diffusing in $\mathrm H$, say). Because of the high thermal diffusivity mediated by photons, $\mathrm{Pr}$ and $\mathrm{Le}$ are very small numbers, some 8-10 orders of magnitude less than in geophysical cases. 

{Such small parameter values} cannot be covered realistically in numerical simulations. If $\tau_{\rm c}$ is a typical convective time scale (as estimated from the superadiabaticity and pressure scale height in the semiconvective zone), very small length scales, of the order $(\tau_{\rm c}\kappa_{\rm S})^{1/2}$ would have to be resolved to represent the interaction between flow and diffusion. This is not computationally possible at present even in 2-dimensional simulations. 

Translation from the numerically accessible parameter values to an astrophysically relevant parameter range therefore requires scaling of the results over the orders of magnitude in-between. A valid extrapolation cannot be found by mere intuitive inspection of numerical results, since these are too far from the target regime. On their own, the applicability of numerical results to the astrophysical case is bound to remain diffuse, or applicability  as an explicit goal has to be given up altogether.

Some physical understanding that includes this asymptotic regime {is necessary}. A theory that accomplishes this is given in the companion paper (S13). This is made possible by making explicit use of the observed separation into layers of convective overturning between stable diffusive interfaces. 

An important difference between the astrophysical and geophysical cases is the equation of state. For the  thin layering expected, sound travel times are very short compared with convective time scales. As a consequence, the fluid in a semiconvective zone behaves as nearly {incompressible}.  A Boussinesq approximation can then be used for the calculations, provided a small complication is properly taken into account.  Whereas in the incompressible case the convective and diffusive heat fluxes are both governed by the temperature gradient, convection and radiative diffusion are governed by different quantities in the compressible case (entropy gradient and temperature gradient). This affects in particular the definition of the Nusselt number as a measure of the efficiency of heat transport. An exact translation between these cases is possible (Massaguer and Zahn 1980, see below in \ref{Bouss}).

A second difference concerns the status of the heat flux in the formulation of the problem. {The effect of semiconvective mixing on the star's structure in evolution calculations is found to be small during the semiconvective phase itself, somewhat independent of the way semiconvection is approximated. (This is in part because of the limited extent of a semiconvective zone). Its effect on the radial profile of elemental composition, however, is of lasting importance for later evolutionary stages (see\ Langer et al. 1989)}.

The consequence {is that}, in contrast with laboratory and geophysical situations, in a stellar model the {heat flux} $F$, rather than the temperature gradient can be considered as known. Since the radiative contribution $F_{\rm r}$ to the heat flux is known to good approximation from the thermal structure of the star, the convective heat flux $F_{\rm c}=F-F_{\rm r}$ transported by semiconvection is also known. The efficiency of convection: i.e. how close {the} mean thermal gradient is to the adiabatic gradient, follows from the imposed heat flux (instead of the other way around as in a laboratory experiment). {We return to this distinction in Sect.\  6.4.}

\subsection{Size of the parameter space}
\label{parsp}

The fluid is described by the thermodynamic variables defining its local state (e.g. temperature and density) and the material functions (e.g. pressure, diffusivities, viscosity). In addition the gradients of the thermodynamic variables with depth, and the acceleration of gravity are relevant for the properties of the flow. Taken together, these quantities form a large parameter space, and it might be concluded that realistic numerical simulations of semiconvection would have to be done for individual zones in individual stellar models. 

The equations of fluid dynamics have symmetries, however, so that the independent degrees of freedom are far fewer. They can be represented by five dimensionless parameters: a Rayleigh number Ra,  the layer thickness $d$ in units of the pressure scale height $H$, the Prandtl number, the Lewis number,  and a density ratio $R_{\rho}$ which measures the ratio of the stabilizing (solute) gradient to the destabilizing thermal gradient. The behavior of semiconvection at some point in a star can be defined in terms of these parameters.

The Boussinesq approximation corresponds to the limit $\epsilon=d/H\ll 1$.  The pressure scale height disappears as parameter in this limit; all dependence on $d$ is subsumed in Ra. This reduces the number of parameters of the problem to four.

By a fortunate coincidence, it turns out  that  as long as the Prandtl number is less than unity, the results are effectively independent of $\mathrm{Pr}$. This further reduces the number of independent parameters to only three. Since measurement of the mixing rate in each individual case does not require a very expensive simulation, this allows a significant volume of parameter space to be covered, and a comparison with the model predictions in S13 to be made.

\section{The layered state}
\label{layerstate}

\subsection{Layer formation, layer thickness}
\label{layform}
The formation of {a} layer from an initially smooth and static gradient starts with a well-known oscillatory instability, with oscillation periods of order the buoyancy period of the stratification. This is followed by nonlinear development into an overturning flow. The transition can be observed in numerical simulations (Merryfield 1995, Rosenblum et al. 2011, Sect. \ref{initial} below), and in very carefully designed laboratory experiments, but is not seen in geophysical cases. An unperturbed smooth initial gradient in solute and temperature is an artificial case that is unlikely to be realized in nature. In addition, the intermediate oscillatory state lasts for only a finite number of oscillation periods (5--10 in the results below), a very short time scale compared with the life time of the double diffusive layers. In the case of lake Kivu, for example, buoyancy periods are in the range 5--30 min, the life time of individual layers a few months.

In laboratory and kitchen table experiments {the layers are initially very thin}. This is understood from the energy argument above. The layers slowly merge, either by fading of contrast between neighboring layers, or the vertical migration of interfaces towards adjacent layers. The details of this process have not been studied much {(but see McDougall 1981, Young \& Rosner 2000, Ross \& Lavery 2009)}. A plausible estimate of the rate of growth of the layer thickness can be given in terms of the effective solute diffusivity of the system, however (S13, and Sect. \ref{stars}). 

In the simulations reported here, the evolution of layer thickness is not included. The thickness ($d$) is therefore treated as a free parameter of the problem. It enters through the Rayleigh number (in the Boussinesq limit, see \ref{parsp} above).  In the astrophysical application of imposed heat flux, however, the resulting mixing rate is effectively independent of the layer thickness (S13 and Sect. \ref{stars} below). This is a fortunate circumstance, since following the evolution of a stack of layers through its merging processes would require far lengthier simulations {(for an example see Young \& Rosner 2000)}.

\subsection{Structure of a layer}

An individual layer in the double diffusive staircase consists of a zone of overturning convection, separated from its neighbors by stably stratified stagnant zones. In a stagnant zone the transport both of heat and solute takes place by diffusion, hence the profiles of $S$ and $T$ are steep and approximately linear with depth in this zone. In the overturning zone the gradients are shallow, except in the thin boundary layers at its interface with the stagnant zone. The thickness $d_\mr{s}$ of the stagnant zone is usually only a small fraction of the layer thickness $d$, but can be as high as 20\% close to marginal conditions for layer formation (for a discussion of its dependence on parameters see S13).

Apart {from} deformations by internal gravity waves in the stagnant zone,  both the vertical and the horizontal velocity components vanish at its interface with the overturning zone. Apart from the small amount of solute carried by the flow, the overturning zone thus behaves essentially like laboratory convection in a box, the interfaces with the stagnant zones acting like the top and bottom plates, with viscosity enforcing no-slip conditions.Because of the periodic boundary conditions in the horizontal direction, free slip conditions would allow the stagnant zones as a whole to start moving sideways. To the overturning flow, the stagnant zones would still present internal no-slip interfaces, however. We have chosen the no-slip conditions since the freedom of such large-scale flows is somewhat of an artefact of the periodic boundary conditions in the horizontal direction. In layers of realistically large horizontal extent they would probably not happen.
This picture has already been put forward early on in the interpretation of laboratory and geophysical observations (Shirtcliffe 1967, Linden \& Shirtcliffe 1978) and of the results from numerical simulations such as those reported below. It has become the standard interpretation used for deducing heat and solute fluxes in thermohaline staircases in geophysics (e.g. Padman \& Dillon 1987, Turner \& Stommel 1964, Schmid et al. 2010).

\subsection{The overturning zone}

\label{overt}
At the boundary with the stagnant zone the overturning flow has three nested boundary layers: for temperature, solute, and flow speed; the thermal, solute and viscous boundary layers respectively. Their thicknesses are determined by the thermal diffusivity $\kt$, solute diffusivity $\ks$, and (kinematic) viscosity $\nu$. The highest diffusivity (thermal) has the widest boundary layer. For astrophysical conditions, $\kt\gg\ks,\nu$, so the solute and viscous sublayers are  thin compared with the thermal boundary layer. 

The horizontal velocity vanishes in the stagnant zone (apart from internal wave modes), so it imposes no-slip conditions on the overturning flow. Since the viscous boundary layer is thin compared with the thermal boundary layer, however, this boundary condition has little effect on turnover times and the resulting heat flux; it becomes noticeable only at low Rayleigh numbers. 

The need to properly resolve the narrowest of the boundary layers in a numerical simulation determines the number of grid points needed in the vertical direction. In the astrophysical case, viscosity is typically of the same order or somewhat larger than solute diffusivity, so the required resolution is set by the solute boundary layer  (cf. Sect.  \ref{setup}).

\subsection{Heat transport}
The flow in the overturning zone is driven entirely by boundary layers at the top and bottom steps of the layer, much like in laboratory convection in a box (Niemela et al. 2000). Except for these thin boundary layers, the (horizontal averages of) entropy and composition are almost uniform inside the layers. Under the conditions in a stellar interior, the thermal diffusivity is much larger than the diffusivity of the solute (Helium in Hydrogen, say). The thickness of the solute boundary layers is then much smaller than the thermal boundary layers, and the amount of solute in transit across the layer vanishes in this limit. The convective {flux} is then almost the same as in the absence of a stabilizing solute (cf. Schmitt 1994). Under these asymptotic conditions, the dependence of heat flux on Rayleigh number can be taken from a simple estimate, as was done in S92, or a result from laboratory measurements of convection can be used (e.g.\  Niemela et al. 2000). With the convective flow thus known, the flux of solute can be calculated as well. 

Convection experiments in gaseous Helium at temperatures just above the critical point by Castaing et al. 1989 yielded the fit
\begin{equation}
{\rm Nu_T}=0.23\ {\rm Ra}^{0.282}. 
\label{NuTCastaing}
\end{equation}
over the range  $10^7<\mathrm{Ra}<10^{14}$.
More recent measurements by Niemela et al. 2000 using superfluid $^3$He, over the remarkable range $10^6 \leq {\rm Ra} \leq 10^{17}$ gave a marginally different result:
\begin{equation}
{\rm Nu_T}=0.124\, {\rm Ra}^{0.309}.
\label{NuTNiemela}
\end{equation}
This can be compared with the estimate based on a 2-dimensional argument in S92:
\beq \nut \approx 0.5\, \mathrm{Ra}_*^{0.25}.\label{s92} \eeq
Since the Prandtl number in the laboratory experiments (of order 0.7) is not far from unity, Ra in (\ref{NuTCastaing}, \ref{NuTNiemela}) can be identified with \rast\ in (\ref{s92}).  Expression (\ref{s92}) then agrees within a factor 2 with (\ref{NuTNiemela}), up to $\mathrm{Ra}_*=10^{15}$.   

The expressions above are of the form ${\rm Nu_T}=a\, {\rm Ra}^b$. They are fits for large Ra; to better cover lower Rayleigh numbers as well, we modify the expression slightly:
\beq \nut-1=a\, (\mr{Ra_*-Ra_{*\mr{c}} })^b,\label{pow} \eeq
where Ra$_{*\mr{c}}=\mr{Pr\, Ra_c}$, and $\mr{Ra_c}\approx 1400$ the critical Rayleigh number for convection with no-slip boundary conditions. {With this change the Nusselt number approaches the diffusive value of 1 as Ra approaches the critical value.}

\subsection{Solute transport}
\label{sol}
The transport of solute is governed by the diffusion from the overturning flow into the steep gradient in the stagnant zone. This happens in a narrow boundary layer of the flow. Its thickness $\delta_\mr{s}$ is of the order $\delta_\mr{S}=(\ks\tau)^{1/2}$, where $\tau$ is the time the flow travels in contact with the boundary before descending/ascending back into the interior. In the same way, the thermal boundary layer has thickness  $\delta_\mr{T}=(\kt\tau)^{1/2}$. The fluxes of solute and heat carried in these boundary layers also determines the flux across the overturning zone as a whole. This predicts that the solute and heat fluxes $F_\mr{S}$, $F_\mr{T}$ are related by
 
\beq F_\mr{S}/F_\mr{T}\sim(\ks/\kt)^{1/2}. \label{squarr}\eeq 
This is the well-known relation derived in various ways for double-diffusive convection (cf. Turner 1985).  In a more accurate form it appears in the model of S13 and in the numerical experiments reported below.

The amount of solute transported across the overturning zone is limited by the fact that convective plumes develop only from material with a net buoyancy of the unstable sign. This limits the density contrast of the stable solute that can be carried to a fraction $1/R_\rho$ of the density contrast across the layer as a whole (Schmitt 1994, S92, S13). Solute with a higher density contrast is not mixed into the convective flow: it remains in the stagnant zone. The solute flux carried by convection in the overturning zone has to match the diffusive flux in the stagnant zone, and idem for the heat flux. In S13 it was shown that these conditions, together with expression (\ref{pow}) form a complete model, determining the effective transport coefficients as well as the thickness of the stagnant zone, as functions of the three parameters of the Boussinesq problem, R$_\rho$, \rast, Le.

In the estimates above the amount of solute transported is given by the amount flowing through the solute boundary layer of thickness $\delta_\mr{s}$. This assumes that the interaction with the stagnant zone is determined entirely by diffusion of the solute. This is a minimum: the stagnant zone as a whole is stable in the sense of the Richardson condition (one verifies that this is equivalent to the fact that R$_\rho>1$), but near its boundary with the overturning zone the density gradient is lower, and a certain degree of mixing by shear instabilities is to be expected. This will increase the amount of solute that is of sufficiently low buoyancy to be carried with the flow. As argued in S13,  this additional amount scales with $\delta_\mr{s}$, so that its net effect is equivalent to an increase of  $\delta_\mr{s}$ by a factor $q$, of order unity, increasing the solute flux by the same factor.  The predicted relation between solute and thermal Nusselt numbers derived in S13, with this erosion factor included is
\beq \nus-1={q\over \mr{Le}^{1/2}R_\rho}\,(\nut-1)\quad (R_\rho<\mr{Le}^{-1/2}). \label{eros}\eeq
The value of $q$ will be a fitting parameter in the quantitative comparison with the numerical results. {[The value $R_\rho=\mr{Le}^{-1/2}$ is the largest for which the model predicts existence of a layered state at any Rayleigh number, see also Sect.\  \ref{max} below.]}

\subsection{Model summary}
The model used for comparison with the numerical simulations is defined uniquely by the ingredients described above. It makes use of only minimal assumptions about of a double diffusive layer:  (a) convection in the overturning zone of a layer is described by a fit to laboratory measurements, (b) transport in the stagnant zone is by diffusion only, and (c) the fluxes of solute and heat are related by the buoyancy limit in the overturning zone (S13 eq.\ 17). These together determine the thickness of the stagnant zone and the effective transport coefficients of heat and solute through the layer as functions of the parameters of the problem, Le, Ra and $R_\rho$. The behavior of the model is discussed in more detail in S13. 

\section{Numerical simulations}
All equations were calculated on a 2D rectilinear Cartesian grid in terms of finite differences. {The horizontal coordinate is labeled with $x$, the vertical coordinate with $z$, where $z=0$ is the bottom and $z=1$ the top boundary.} The set of equations are implemented into the ANTARES software framework (Muthsam et al.\ 2010). Advective currents are treated by a weighted essentially non-oscillatory finite volume scheme in fifth order (Shu \& Osher 1988), the physical diffusion is handled by a fourth-order finite difference discretization. A second order total variation diminishing scheme is chosen as time integrator. To avoid odd-even decoupling, a MAC grid, which locates vector variables at cell faces and scalar variables at cell centres, is used. 

For a detailed description of the numerical solution of the binary mixture equations presented here see Zaussinger 2011.

\subsection{Boussinesq approximation for a binary mixture}
In the Boussinesq approximation the continuity equation is reduced to that of an incompressible fluid,

\begin{equation}
\frac{{\mathrm d} \rho}{{\mathrm d} t} =  0, \label{BA1}
\end{equation}
such that in the equation of motion the density  is taken to be a constant $\rho_0$,

\begin{equation}
\frac{{\mathrm d} \vec u}{{\mathrm d}t} = -{1\over\rho_0}\nabla P + \left ( - \alpha_{\rm T}\Theta + \alpha_{\rm S}S \right ) {\vec g} + \nabla \cdot (\nu \nabla \vec u),  \label{BA2}
\end{equation}
where the `expansion coefficients' $\alpha_{\rm T}, \alpha_{\rm S}$ describe the density effects   of variations $\Theta$, $S$ in temperature and solute, assumed small compared with the mean temperature $\bar\Theta$ and salinity $\bar S$. They are given by advection-diffusion equations: 

\begin{equation}
\frac{{\mathrm d} \Theta}{{\mathrm d}t }= -\vec u \cdot \nabla \bar{\Theta} +  \nabla \cdot  (\kappa_{\rm T} \nabla \Theta),  \label{BA3}
\end{equation}

\begin{equation}
\frac{{\mathrm d} S}{{\mathrm d}t}= -\vec u \cdot \nabla \bar S + \nabla \cdot  (\kappa_{\rm S} \nabla S),  \label{BA4}
\end{equation}
where $\nu$ is the kinematic viscosity. The mean gradients $\nabla\bar\Theta$, $\nabla\bar S$ can be expressed more usefully in terms of the buoyancy frequencies:
\beq N_{\rm T}^2=\alpha_{\rm T}\,{\bf g}\cdot\nabla\bar\Theta,\label{NTB}\eeq
\beq N_{\rm S}^2=\alpha_{\rm S}\,{\bf g}\cdot\nabla\bar S,\label{NSB}\eeq
The density ratio $R_\rho$ is then defined as in the compressible case, eq.\ (\ref{rrho}).

The numerical algorithm solving this set of equations is based on a semi-implicit scheme. Intermediate values for the velocity field $\vec u^*$ are calculated explicitly from the equations of motion. By the nature of the incompressible equations, the pressure update is done implicitly, by solving a Poisson equation:

\begin{equation}
\Delta P = \frac{\rho_0}{\Delta t} (\nabla \cdot \vec u^*)  \label{Poisson}
\end{equation}

The resulting pressure $\rm P$ leads to the required divergence free velocity field at the new time step $n+1$.

\begin{equation}
\vec u^{n+1} = \vec u^* - \frac{\Delta t}{\rho_0} \nabla P.  \label{update}
\end{equation}

\subsection{Compressible fluid equations for a binary mixture}
Verification of the Boussinesq results  has been done with simulations of the fully explicit compressible fluid equations. The fluid is assumed to be an ideal gas, which is a good approximation to a binary gas mixture of our interest. These are the continuity equation

\begin{equation}
\frac{\partial \rho}{\partial t} + \nabla \cdot (\rho \vec u )=0
\end{equation}

the partial density equation
\begin{equation}
\frac{\partial (\rho c)}{\partial  t} + \nabla \cdot (\rho  c  \vec u )= \nabla \cdot (\rho \kappa_c \nabla c)
\end{equation}
 
the momentum equation
\begin{equation}
\frac{\partial (\rho \vec u)}{\partial t} + \nabla \cdot (\rho \vec u\vec u + P \underline{\rm I} )= \rho  {\bf g} + \nabla \cdot \tau
\end{equation}
 
the total energy equation
\begin{equation}
\frac{\partial \rho E}{\partial t} + \nabla \cdot [ \vec u( {\rho E + P}) ] = \nabla \cdot (K_{\rm h} \nabla T) + \nabla \cdot (\vec u \tau) +  \rho \vec  g \vec u
\end{equation}

and the equation of state

\begin{equation}
P = \frac{\mathcal{R} \rho T}{\mu (c)}
\end{equation}

where $c$ is the solute mass fraction,$E=\frac{1}{2}|\vec u|^2 + e$ is the total specific energy in units of energy per mass, $e$ is the specific internal energy, $K_{\rm h}$ is the heat conduction coefficient, $\kappa_c$ is the solute diffusion coefficient, $\tau$ is the viscous stress tensor, {$\bf g$ the gravitational acceleration, \underline{\rm I} the unit tensor}, $\mu = \mu(c)$ the mean molecular weight, and $\mathcal{R}$ the gas constant.   

\subsection{Units, boundary and initial conditions}
\label{bc}

As unit of length we use, for the Boussinesq cases the layer thickness $d$, for the compressible calculations the pressure scale height $H$.  A nominal convective turnover time is used as unit of time. As units of temperature (potential temperature $\Theta$ in the compressible case) we use $1/\alpha_{\rm T}$, for solute concentration $1/\alpha_{\rm S}$. The density ratio then becomes $R_\rho=\nabla \bar S/\nabla \bar T$. The Boussinesq equations are invariant to arbitrary additive constants in temperature and solute. We set these such that $T$ and $S$ are zero at the top boundary.

Because of the symmetries of the problem, there are fewer independent parameters than physical variables describing it. Hence some of the physical quantities appearing in the problem can be set to unity. We choose for these: the temperature difference between top and bottom of the layer, the density at the bottom of the layer and the acceleration of gravity. The Rayleigh number Ra$_*$ is then controlled through the thermal diffusivity $\kappa_{\rm T}$, the solute difference between top and bottom through the density ratio $R_\rho$, and the solute diffusivity through the Lewis number $\mathrm{Le}$. 

Most of the calculations were done in a box simulating a single layer from the double-diffusive staircase, so the top and bottom boundaries coincide with the steps between layers. This ignores the distortions of the interfaces by surface waves, but since the essence of the double layering phenomenon is that the transport across the interface is by diffusion, this is not expected to make a big difference. To check that this is indeed the case, a smaller set of simulations was done in which a step is present inside the volume (Sect.\  \ref{double}). The vertical boundary conditions are thus taken to be impermeable and stress-free. {}In the horizontal direction periodic conditions are used.   
\begin{eqnarray}
u_z&= 0 \quad ~~(z = & 0, 1) \\
\frac{\partial u_x}{\partial z}&=0 \quad ~~(z =& 0, 1) \\
S &=R_\rho  ~~~(z =& 0) \\
T &=1 ~~~~~(z =& 0) \\
S&= 0 ~~~~~ (z = &1) \\
 T&= 0 ~~~~~ (z =& 1)
\end{eqnarray}      
As initial condition the stratification of temperature and solute was taken to be horizontally uniform with either a linear gradient between the values at top and bottom (the `linear' case below), or something approximating the boundary layer structure expected of the final state (the `step' case). Small initial random perturbations are applied on the solute field. 

The numerical algorithm for setting up the initial conditions for the compressible fluid equations is an extension of the procedure presented in Muthsam et al.\ (1995, 1999).

\subsection{Numerical setup}
\label{setup}
Being the thinnest boundary layer in the problem, $\delta_{\rm S}$ determines the numerical resolution needed near the boundaries. Since the fine structure in the interior of the layer largely consist of boundary layers `peeled off' from the boundaries, the same resolution is needed in the interior of the layer as well, and there is no need or justification for using non-uniform grids.

The expected solute boundary thickness for  Ra$_*=1.6 \times 10^5$ and a Lewis number of $\rm Le = 0.1$ is $1.6\%$ of the layer thickness $d$. With the high-order spatial discretization used, a nominal resolution of $n_{\rm BL}=3$ points is needed across the critical structures that must be captured. The transport of solute across the layer is determined by the vertical structure of the boundary layers. Resolving these with $n_{\rm BL}=3$ translates to $n_z=200$ points needed in the vertical direction for this case. 

To test the actual convergence of the results with respect to resolution, a series of simulations with $n_z=100$, $200$, $300$ and $600$ has been run for Ra$_*=1.6 \times 10^5$, $\rm Le = 0.1$, $R_\rho=1.15$,  see Table \ref{table:res}. Convergence to 15\% is reached already at $n_z=100$. For the lowest Lewis number used in the results reported below in Table 2, Le=0.01, the number of grid points needed would be $\sqrt 10$ times higher. The number of points across the height of the box used in these simulations, $n_z=300$, is deemed sufficient for the level of accuracy we are aiming for.
\begin{tiny}
\begin{table}[h]
\centering
\begin{tabular}{c | c | c }
$n_z$ & \nust & \nutt \\
\hline
$100$ & $26$ & $9.5$ \\
$200$ & $27$ & $10$ \\
$300$ & $30$ & $10$ \\
$600$ & $30$ & $10$ \\
\hline
\end{tabular}
\caption{A test series shows convergence of the Nusselt numbers with respect to resolution (number of grid points $n_z$) across the height of the box.}
\label{table:res}
\end{table}
\end{tiny}

Most of the calculations were done with a horizontal-to-vertical aspect ratio of 2:1. A few tests with different ratios showed that this choice does not affect the measured Nusselt numbers significantly (Sect.\  \ref{aspect})

\section{Results of numerical simulations}

Over a wide range in parameter values about $100$ numerical simulations have been done in the Boussinesq approximation and about $20$ simulations  {have been} performed with the fully compressible code.  Only the  cases closest to the parameter range of astrophysical interest (about 60) are reported here (Figs. \ref{floplots}, \ref{DepPrNuS}, Table 1).  

\begin{figure}
\begin{center}
\includegraphics[width=0.9\hsize]{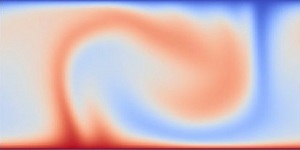}\\
\vspace{0.4\baselineskip}
\includegraphics[width=0.9\hsize]{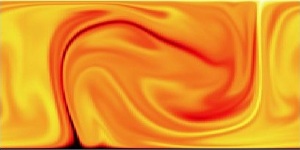}
\caption{Flow structure in a double diffusive layer. The temperature field (top) is more diffuse than the solute (`Helium') concentration, as a result of the high thermal diffusivity. }
\label{sample}
\end{center}
\end{figure}

An example of the flow structure is shown in Fig.\ \ref{sample}, with $\rm Pr=0.1$, Ra$_*=5\times10^5$, $\rm Le=0.01$, R$_\rho$=1.15. It shows the key characteristic of double diffusive convection:  the  boundary layers and the plumes of the solute are narrower than those of temperature, on account of the low solute diffusivity. 

\subsection{Dependence on {\rm \rast} and $R_\rho$}

\label{lerho}
\begin{figure*}
\begin{center}  
\includegraphics[width=0.8\textwidth]{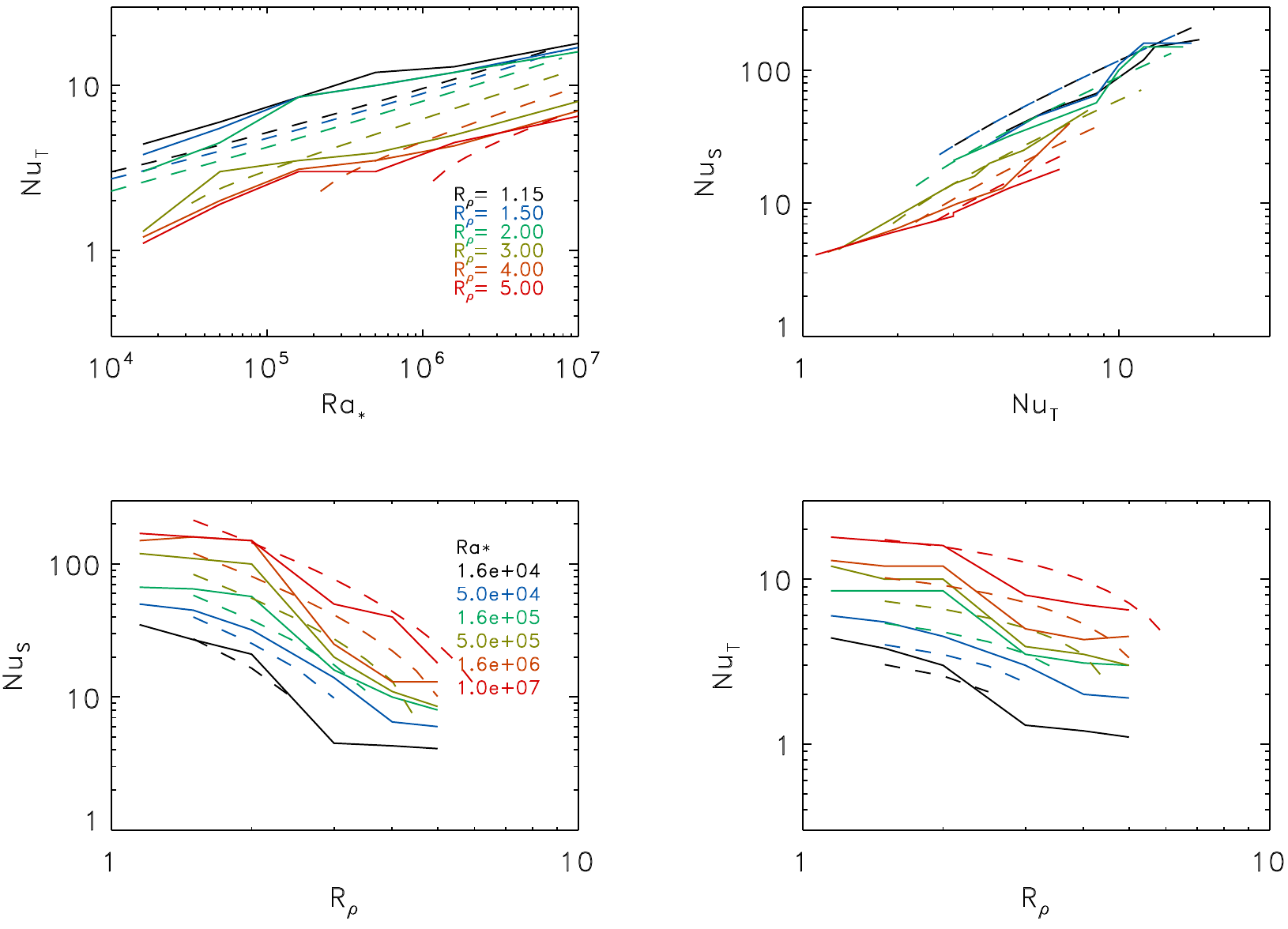}
\caption{Dependence of the Nusselt numbers on density ratio $R_\rho$ and Ra$_*$, for Le=0.01, Pr=0.1. Solid: numerical results, dashed: model predictions for erosion factor $q=1.5$.}
\label{floplots}
\end{center}
\end{figure*}

 Fig.\ \ref{floplots} shows the dependence of the measured Nusselt numbers \nutt and \nust on the parameters \rast\ and R$_\rho$, for the case Le$=0.01$, Pr=0.1. The results show some fluctuation, as expected from the limited number of overturning times for which the simulations were run. This results in uncertainties in the Nusselt numbers of the order of a factor 1.5. For comparison with the theoretical model predictions (dashed lines) the adjustable erosion factor $q$ in eq.\  (\ref{eros}) has been set to $1.5$. Within the scatter, the predictions appear consistent with the numerical results to a factor of 2 or less. In fact, leaving out the erosion factor (i.e. setting $q=1$) does not make the fit {a lot worse}. 
   
%, with parameters Ra (along the curves) and R$_\rho$ (between the curves).
The upper right panel shows Nu$_\mr{s}$ vs. Nu$_\mr{T}$, where each curve represents the results for a fixed density ratio R$_\rho$ and concurrently increasing $\ras$. The approximately linear relation between the two shows that the {ratio} of thermal to solute transport  depends primarily on Lewis number and density ratio, but not much on Ra, as predicted by (\ref{eros}). The actual transport efficiency of both depends of course on the Rayleigh number Ra$_*$ as well (upper left panel).
   
\subsection{Behavior near the maximum density ratio}
\label{max}
\begin{figure}
\begin{center}  
\includegraphics[width=\hsize]{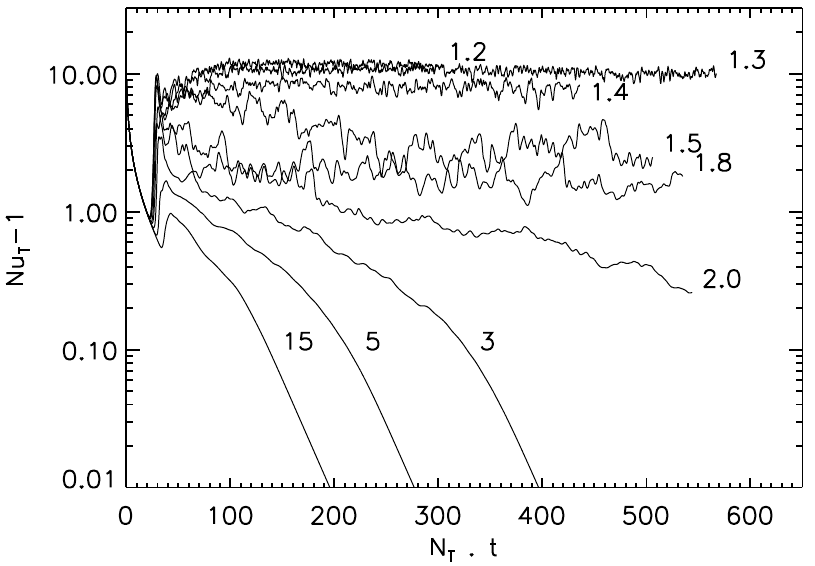}
\caption{Time dependence of the thermal Nusselt number $\nut$ in simulations with Pr=1, Le=0.1, $\ras=10^6$, for density ratios (labeled) near the maximum value for which a layered state is predicted by the theoretical model. Time t in units of the thermal buoyancy time scale $N_{\rm T}^{-1}$.}
\label{tdep}
\end{center}
\end{figure}

The data in Fig.\ \ref{floplots} show how the Nusselt number \nutt  declines with increasing density ratio. {The theoretical analysis in S13 predicts a maximum density ratio $R_{\rho\,{\rm max}}$ for the existence of a (statistically) stationary layered state. It is a slowly increasing function of \rast, asymptotically approaching the value ${\rm Le}^{-1/2}$. To test this a series of simulations was done at $\ras=10^6$, Le$=0.1$. The predicted value of  $R_{\rho\,{\rm max}}$ for this combination is $1.2$ (the asymptotic value would be $1.3$). Fig.\ \ref{tdep} shows the resulting Nusselt numbers as functions of time. }

For $R_\rho=1.2$ and 1.3, the heat flux in the simulations quickly settles to a statistically steady value after a transient due to the initial conditions used. For the three {highest} values of $R_\rho$, the initial state smoothly settles to the purely diffusive state $N_{\rm T}=1$, {the faster and smoother, the higher the density ratio}. Values in-between (1.5 and 1.8) are characterized by large fluctuations superposed on a slow general decline, as if the system {is undecided between settling on an overturning or a static diffusive} state. A precise boundary for the existence of an overturning state is therefore somewhat hard to define: it appears to depend on the length of time over which the flow is followed. 

{The time dependences of the Nusselt number  shown in Fig.\ \ref{tdep} suggest a value $R_{\rho\,{\rm max}}\approx 1.4$, somewhat larger than the theoretical value of 1.2. For density ratios slightly larger than 1.2, it looks at first as if a statistically steady state has been reached, but on a longer time scale a continuing decline of \nutt is observed.}  {The behavior with increasing $R_\rho$ is therefore somewhat gradual, at least for the finite length of time over which the simulations have followed the development of the flow. Interestingly, the distinction between diffusive and overturning final states becomes sharper with increasing length of time over which the simulation is followed. This also affects the results in Fig.\ \ref{floplots}. At the largest density ratios shown there, the numerically measured Nusselt numbers are systematically higher than predicted. This can be attributed to the fact that at these $R_\rho$ the simulations were not run as long as those of Fig.\ \ref{tdep}. }

The numerical results thus confirm the predicted existence of a maximum density ratio for the overturning layered state, but with the added twist of very long settling times when $R_\rho$ is near the maximum value. This also answers a question that was left undecided in S13: the theory does not predict what happens close to the maximum density ratio. The results in Fig.\ \ref{tdep} indicate that the system vacillates between overturning and diffusive states for increasingly long periods of time as the maximum is approached.

The fast settlement towards a diffusive state for high values of $R_\rho$ is not surprising given that the linear stability criterion predicts exactly such a result for $R_\rho > (1+Pr)/(Le+Pr)$ (e.g Veronis 1968, Stevenson 1979). For the parameter values of Fig. 3 this yields $R_\rho\ga 1.8$. The difference between 1.8, where $Nu_T -1$ drops only initially, but then only mildly so if at all, and a value of 2.0, which shows a continuous decay seems consistent with this classical result.

The results also provide an interesting link to Proctor's (1981) analysis.  In this theory the maximum value $R_\rho={\rm Le}^{-1/2}$ appears as a limit for its applicability. Proctor's  analysis therefore does not answer what happens near this critical number, any more than the model in S13 does, but it is gratifying that it appears to have a consistent place in both, as well as in the numerical results.

\subsection{Dependence on $\rm Pr$}
\label{prdep}
\begin{figure}
\begin{center}
\includegraphics[width=0.7\hsize]{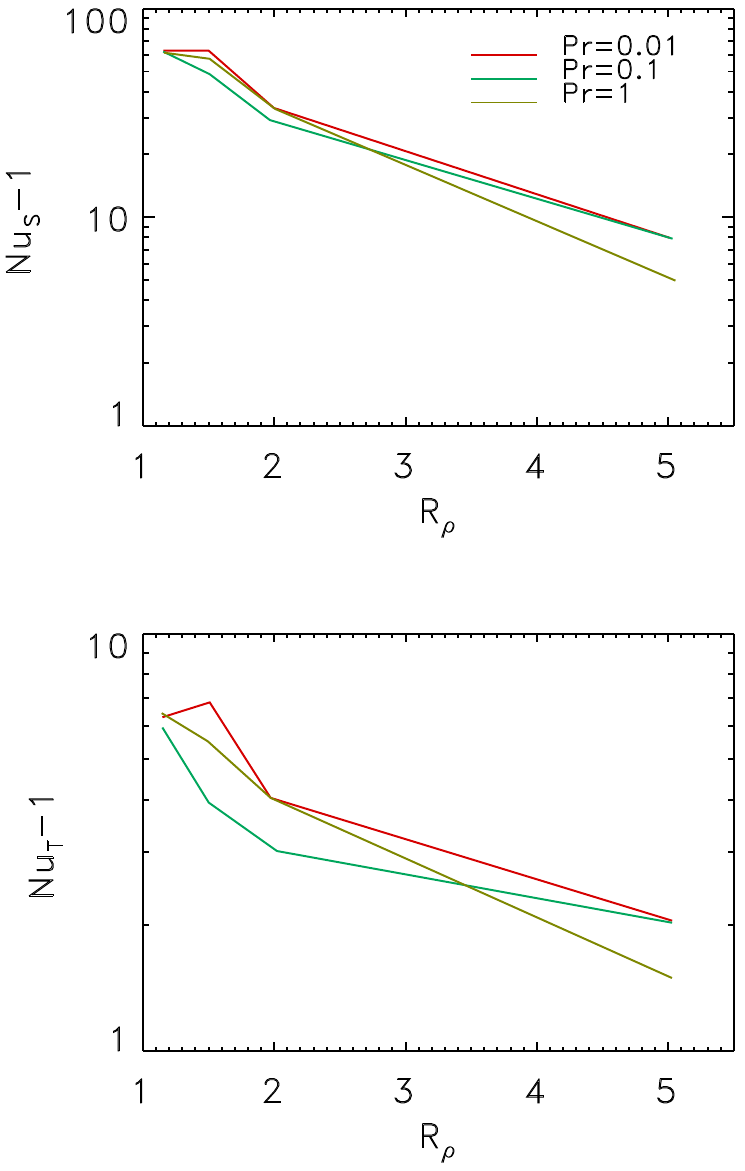}
\label{DepPrNuT}  
\caption{Dependence on Prandtl number and density ratio.  Solute Nusselt number (top) and  thermal Nusselt number  (bottom) for $\rm Pr=1$, $\rm 10^{-1}$ and $\rm 10^{-2}$, with Ra$_*=1.6\,10^5$ and $\mathrm{Le}$ fixed at 0.01.}
\label{DepPrNuS}
\end{center}
\end{figure}

Fig.\ \ref{DepPrNuS}  shows the results of a set of simulations testing the dependence on Prandtl number. The Lewis number is fixed at $\mathrm{Le}=0.01$. The density ratio is varied over the range $1.15<R_\rho<5$, Prandtl number from 0.01 to 1. Within the limited numerical sensitivity due to the stochastic nature of the flow, little systematic dependence on Pr is detectable.  As argued above, a low dependence on Pr was to be expected in the limit $\mathrm{Pr}\downarrow 0$. It appears { that this extends} also to a Prandtl number of order unity.

\subsection{Dependence on aspect ratio}
\label{aspect}
The aspect ratio of $2:1$ used in the results reported above was tested against $5:1$ and $10:1$ at the same spatial resolution. For the reference simulation ($\rm Pr=0.1$, $\rm Le=0.01$, Ra$_*=10^5$, $R_{\rho}=1.15$) we find  Nu$_S=90$ and $\nut=8.5$. By comparison the simulation with $5:1$ results in $\nus=90$ and $\nut=9.0$. The most extended box with an aspect ratio of $10:1$ and a spatial resolution of $1500\times300$ has Nusselt numbers of  $\nus=80$ and $\nut=8.75$.  The aspect ratio thus has no significant influence on the dependences of the fluxes on input parameters in our simulations, within the fluctuations due to  the stochastic nature of the flow.

\subsection{Dependence on initial stratification}  
\label{initial}
\begin{figure}
\begin{center}
\includegraphics[width=0.45\hsize]{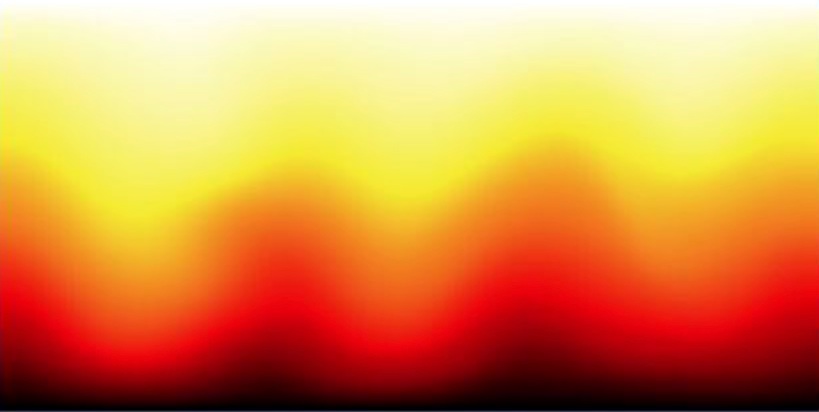}
\includegraphics[width=0.45\hsize]{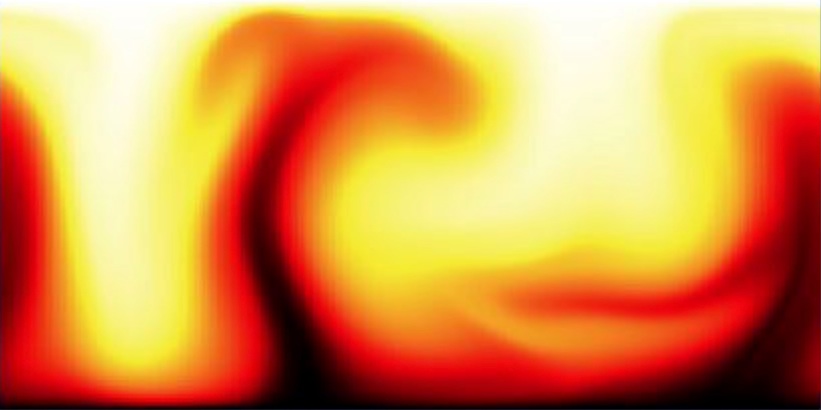}\\
\includegraphics[width=0.45\hsize]{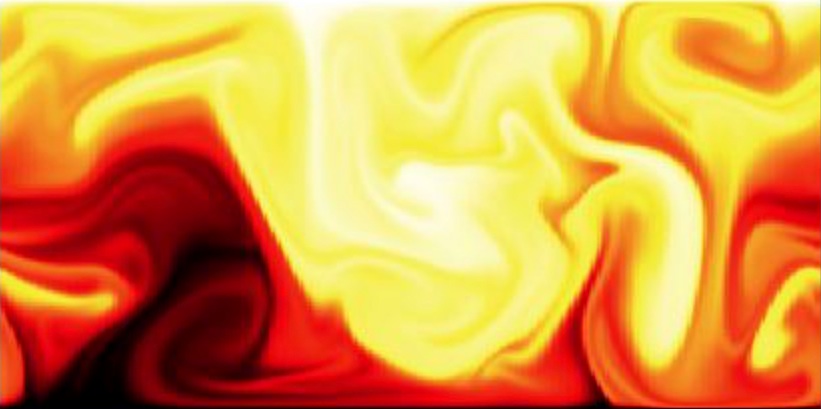}
\includegraphics[width=0.45\hsize]{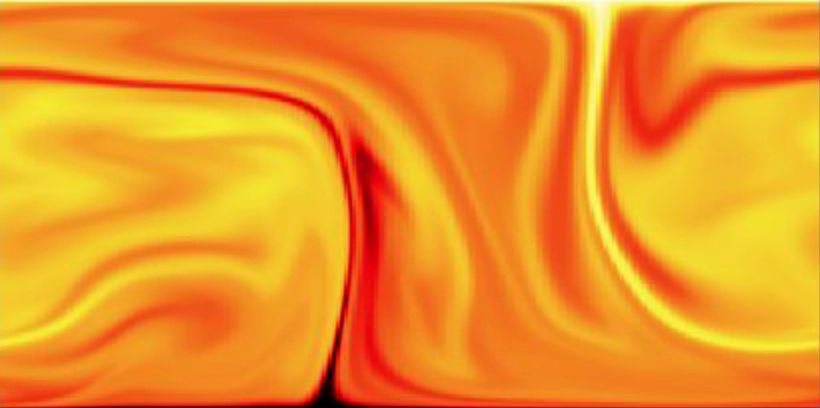}
\caption{Example of the development of an overturning flow from Kato oscillations for $\rm Pr=1$, $\rm Le=10^{-1}$, $R_{\rho}=1.15$ and Ra$_*=1.6\times10^5$ starting from a linear stratification. Time from left to right and top to bottom. {Wave braking occurs after 5 oscillations (Fig.\  5.2). The layer is fully evolved after 10 oscillations (Fig.\  5.4). See also the movie at http://www.mpa-garching.mpg.de/$\sim$henk/movie.avi}}
\label{kato}
\end{center}
\end{figure}

\begin{figure}
\begin{center}
\includegraphics[width=1.02\hsize]{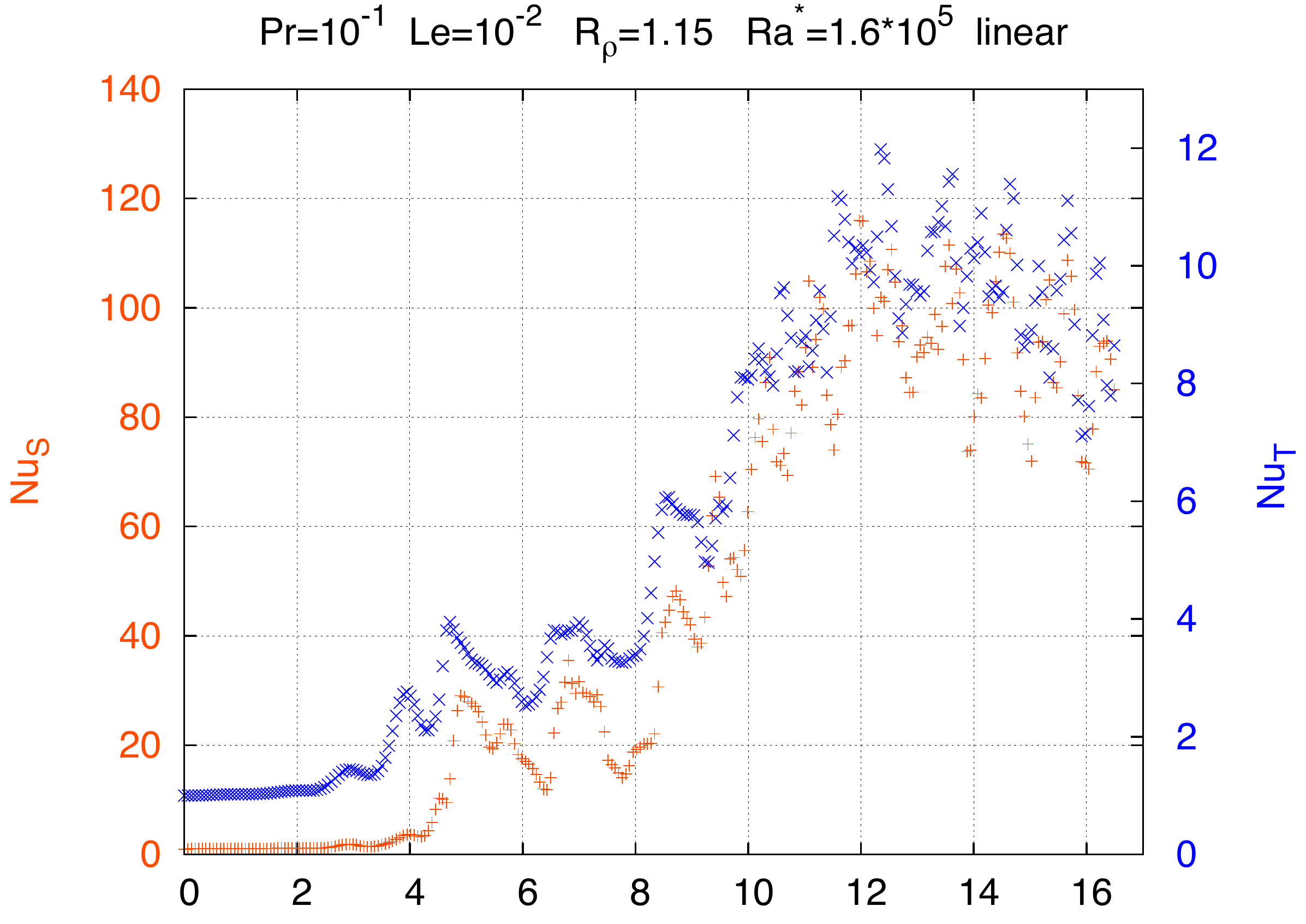}
\includegraphics[width=1.02\hsize]{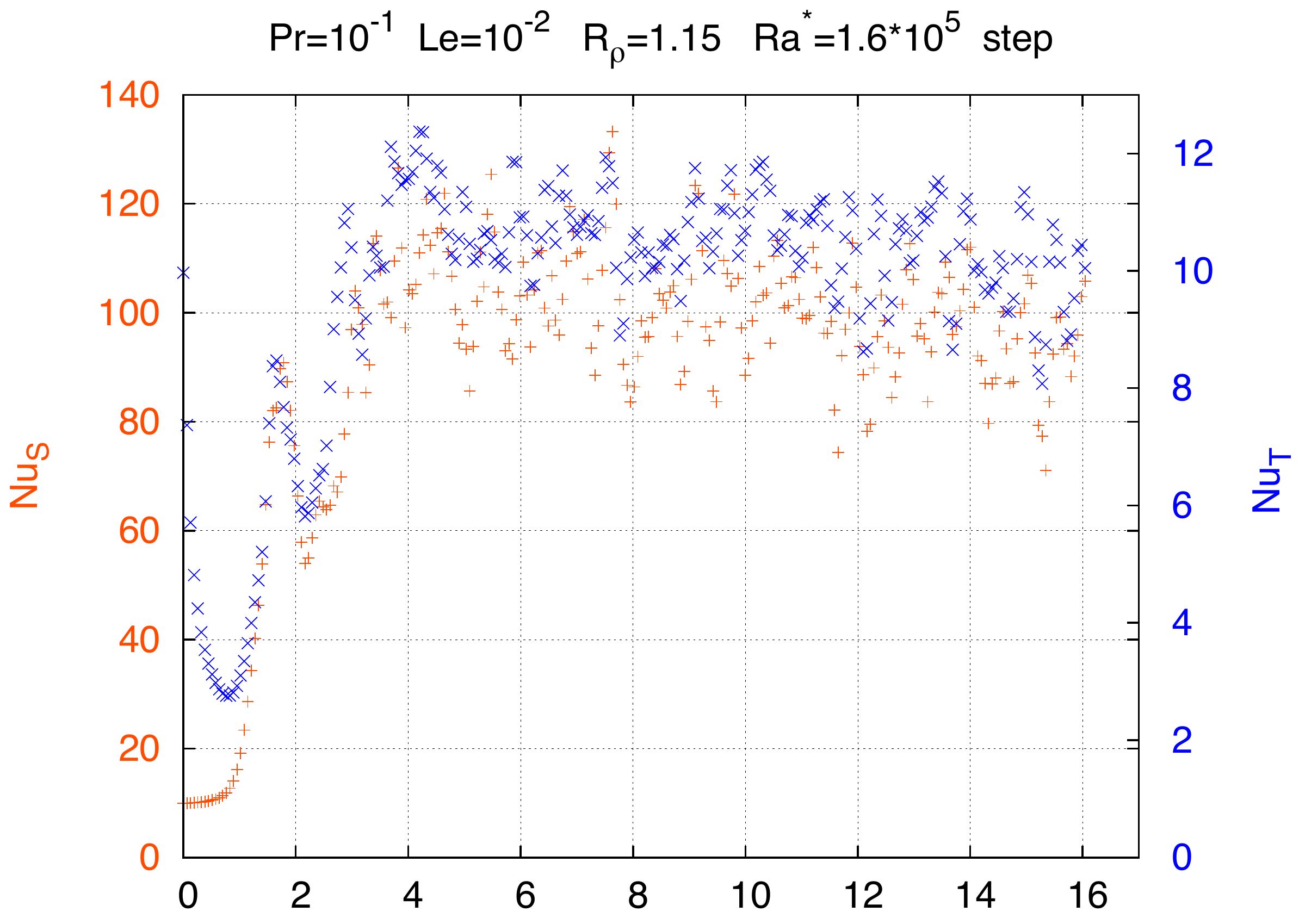}
\caption{Top panel: evolution of the Nusselt number for the linear initial stratification. Convective cells get established from Kato oscillations (cf. Fig.\ \ref{kato}) after about 10 turnover times. Starting the simulation from a step (bottom) saves computing time. At the end of the runs the Nusselt numbers of both simulations are the same within the statistical variations.}
\label{fig:evolution}
\end{center}
\end{figure}

As initial state we used either a constant linear gradient of temperature and solute between top and bottom values (`linear'), or a profile approximating the expected steplike combination of a stagnant and an overturning zone. The linear case shows how the oscillatory phase due to the Kato instability develops into overturning flow, see Figs.\ \ref{kato}, \ref{fig:evolution}. The duration of the initial formation process is mainly determined by $R_{\rho}$ and Ra$_*$. It is of the order of 5--10 oscillation periods of the initial stratification. The end state in both cases is statistically the same. The `step' as initial condition also covers cases that are stable in linear theory because of the subcriticality of the system, and would not develop from a linear initial profile. With this initial state computing time can be saved for small values in Ra$_*$ and high $R_{\rho}$. It was used in most of the results reported above.

\subsection{Comparison with compressible results}
\label{compar}
The compressible simulations  are based on a  5th order weighted ENO scheme. 
Compressible fluids lead to restrictions in time stepping (due to the need to resolve sound waves). The compressible simulations take up to $100$ times longer for the same resolution compared to the incompressible solver. In both regimes a resolution of $300\times300$ points has been set. Therefore only a few tests have been done with the compressible code. The degree of compressibility is governed by the $\epsilon=d/H$ of layer thickness to pressure scale height. In the limit $\epsilon\rightarrow 0$ there is a direct translation between the compressible and the Boussinesq case (Sect.\  \ref{Bouss}). The results of a numerical comparison with $\epsilon=0.1$ is shown in Table 1. The resulting Nusselt numbers do not differ significantly.

Simulations done with $\epsilon=1.0$ behave quite similar to {those} done with $\epsilon=0.1$. The mixing processes do not significantly differ{, at least for $\rm Pr \leq 0.1$ and} as long as the Rayleigh number is high enough, $\mathrm{Ra} >5.0 \times 10^5$. Differences of order 50\% in the Nusselt numbers {were} found, but considering the present level of accuracy (a factor of 2, say), we conclude that the Boussinesq approximation qualitatively gives the right results, even for layer thicknesses approaching a scale height. At Pr of order unity or larger, the effect of viscous damping might become more significant in more strongly stratified cases, however.
\begin{tiny}
\begin{table}[h]
\centering
\begin{tabular}{c | c | c | c | c | c | c | c}
Pr & Le & $R_{\rho}$ & $\ras$ & $\rm Nu_S^B$ & $\rm Nu_T^B$  & $\rm Nu_S^s$ & $\rm Nu_T^s$\\
\hline
\vspace{1\baselineskip}
$10^{-1}$& $10^{-2}$ & $2.0$ & $1.6 \times 10^5$ & $60$  & $8$ & $55$  & $7$  \\
$10^{-1}$& $10^{-2}$ & $1.2$ & $1.6 \times 10^5$ & $110$ & $12$ & $110$  & $11$  \\
$10^{-1}$& $10^{-2}$ & $2.0$ & $1.6 \times 10^6$ & $150$  & $12$ & $130$  & $10$  \\
$10^{-1}$& $10^{-2}$ & $1.2$ & $1.6 \times 10^6$ & $200$  & $16$ & $200$  & $14$ \\
$1.0$ & $10^{-1}$ & $2.0$ & $1.6 \times 10^5$ & $3.5$ & $2$ & $11$  & $1.5$  \\
$1.0$ & $10^{-1}$ & $1.2$ & $1.6 \times 10^5$ & $45$  & $15$ & $17$  & $5$  \\
$1.0$ & $10^{-1}$ & $2.0$ & $1.6 \times 10^6$ & $4$  & $3.5$ & $26$ & $10$  \\
$1.0$ & $10^{-1}$ & $1.2$ & $1.6 \times 10^6$ & $33$  & $11$ & $26$  & $10$  \\
\hline
\end{tabular}
\caption{Comparison of compressible and incompressible simulations. Thermal ($_{\rm T}$) and solute  ($_{\rm S}$) Nusselt numbers from the Boussinesq ($^{\rm B}$) and compressible ($^{\rm s}$) results. Layer thickness $d$ is 0.1 pressure scale height. }
\label{table-eps01-2}
\end{table}
\end{tiny}

\subsection{Multi-layer simulations}
\label{double}

\begin{figure}
\centering
\includegraphics[height=0.8\hsize]{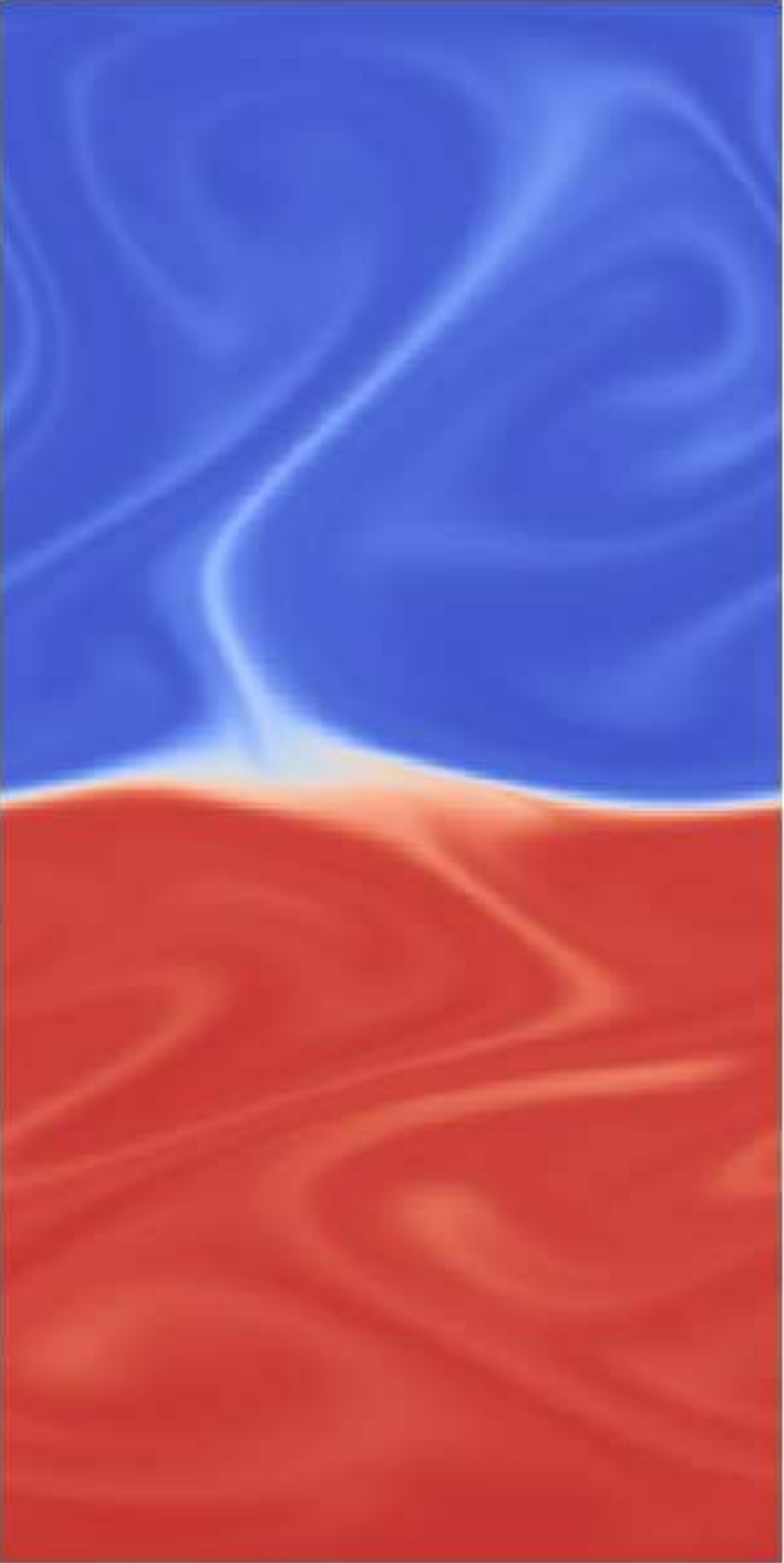}\hfil
\includegraphics[height=0.8\hsize]{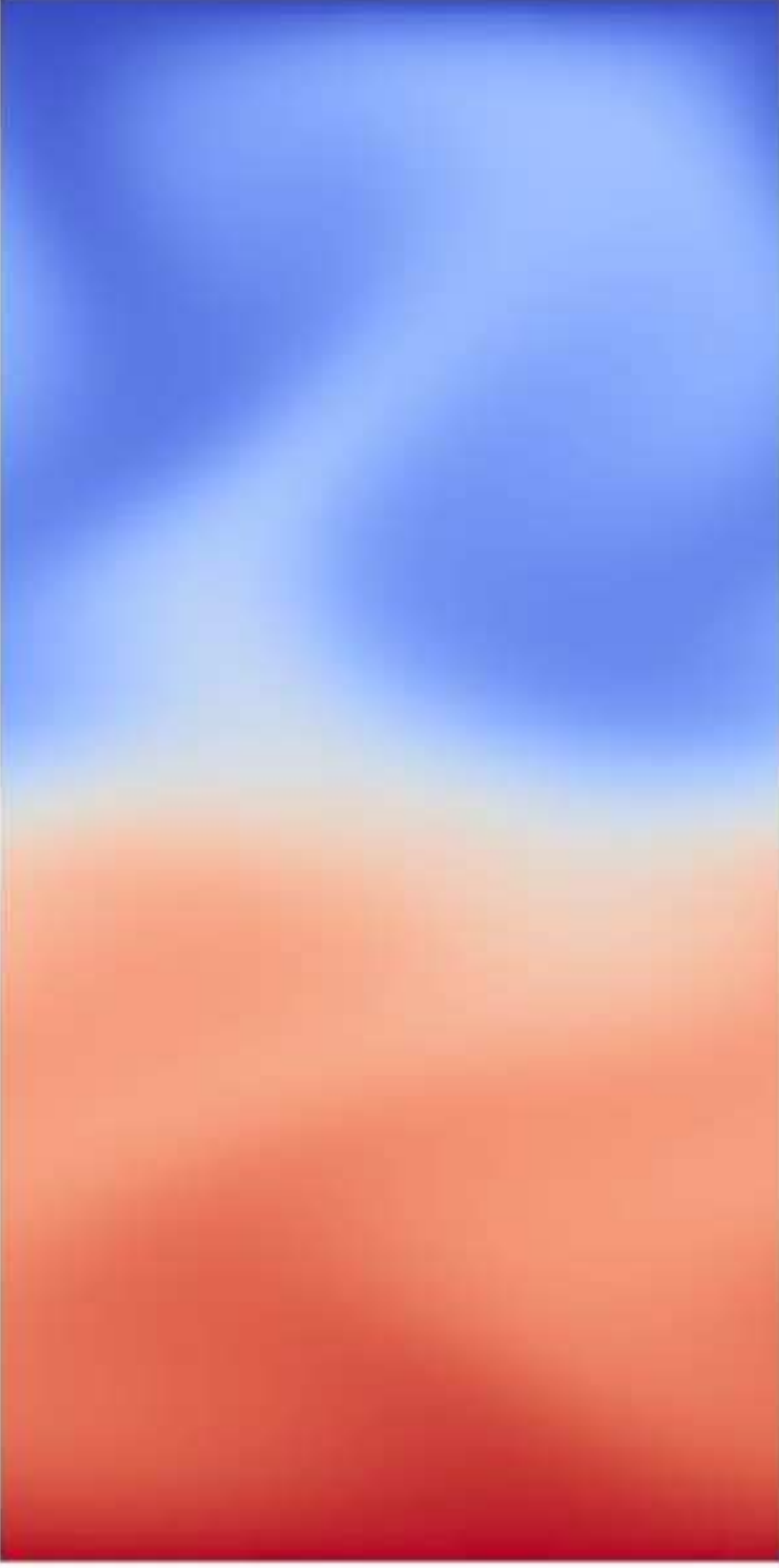}

\caption{Snapshot of a simulation including the free interface between two layers. Left: solute, Right: temperature. Pr=1.0, Le=0.01, $R_\rho$=1.15, Ra$_*=6\,10^5$. See movie at http://www.mpa-garching.mpg.de/$\sim$henk/double\_layer.avi } 
\label{fig:ds1}
\end{figure}

In all of the above we have assumed that the interfaces between the double diffusive layers can be approximated as solid boundaries. To test the reliability of this assumption, a few cases were run where the initial state consisted {of two steps instead of a single one}. In some, though not all of these runs, the division into two layers remained till the end. An example is shown 
in Fig.\ \ref{fig:ds1} for a case with $\rm Pr=1$, $\rm Le=10^{-2}$, $R_{\rho}=1.15$, Ra$_*=6\times 10^5$ and a resolution of $500\times 500$. Note the approximate (anti-)symmetry of the plumes near the interface in the middle, a phenomenon known from laboratory experiments (e.g. Fig. 1 in Turner 1985). It is caused by the continuity of the horizontal velocity across the interface enforced by viscosity. This symmetry is less marked in simulations at lower Prandtl numbers, when the thickness of the viscous boundary layer becomes smaller than the thickness of the stagnant zone.

\begin{figure}
\centering
\includegraphics[width=\hsize]{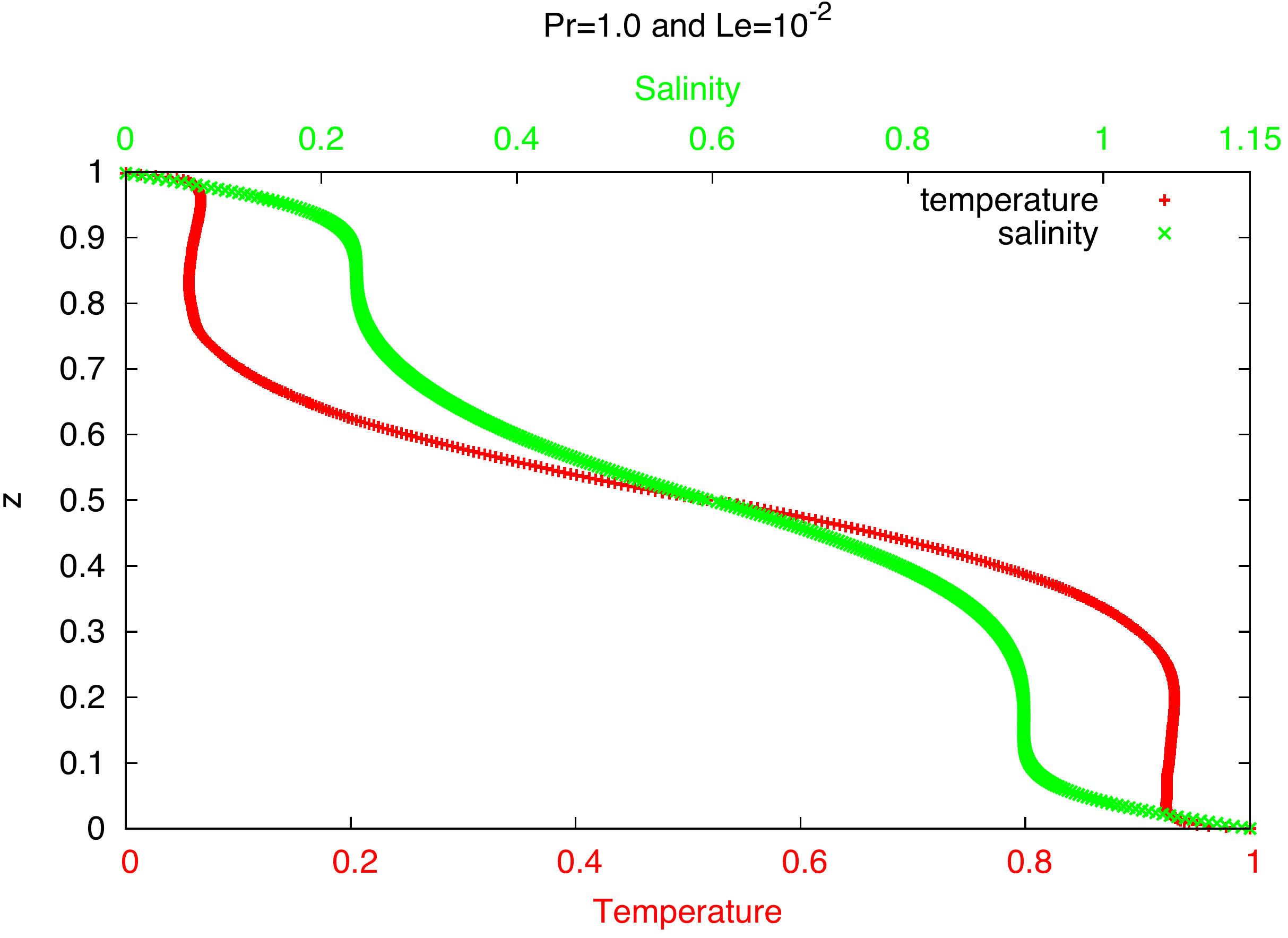}\hfil
\caption{Average temperature and solute profiles with height $z$ in the double-layer simulation of Fig.\ \ref{fig:ds1}.} \label{dsmean}
\end{figure}

Fig.\ \ref{dsmean} shows horizontal and temporal average profiles of temperature and solute with height. The transition in the middle is  broad compared with the boundary layers at top and bottom. Inspection of the time dependent flow shows that this is due to two separate effects. One is the displacements of the interface by surface waves, which smoothen the average gradient without changing the actual fluxes across the interface. In addition there is possibly some real mixing associated with breaking of the surface waves, but it remains localized around the interface between overturning and stagnant zone.

\section{Application to stars}
\label{appli}

\subsection{Boussinesq limit: thin layers} 
\label{Bouss}

If the double-diffusive layering is thin, convective overturning times are long compared with the sound crossing time of the layer thickness, and a Boussinesq (or other `low Mach number' approximation) can be used for a compressible fluid.  By the presence of a stratification in density and temperature, however, a heat flux is present even in a convectively neutral stratification. This requires some care in translating Boussinesq results to the astrophysical case. 

\subsection{Heat flux}
Let $F_{\rm r}$, $_{\rm c}$ be the radiative and hydrodynamic (semiconvective) contributions to the heat flux in the star. The radiative heat flux is proportional to the logarithmic temperature gradient $\nabla=\mr{d}\ln T/\mr{d}\ln P$:
\beq
F_{\rm r}= k\nabla= k\nabla_{\rm a} + k(\nabla-\nabla_{\rm a})\equiv F_{\rm ra}+F_{\rm rs},\label{frad}
\eeq
where $k$ is a constant depending on the local thermodynamic state, $\nabla_\mr{a}$ the adiabatic gradient, and $F_{\rm ra}$, $F_{\rm rs}$ are the contributions to the \underline{r}adiative heat flux of the \underline{a}diabatic and the \underline{s}uperadiabatic parts of the mean temperature gradient{,} respectively. In the Boussinesq model, the contribution $F_{\rm ra}$ is absent: the convective and radiative heat fluxes $F_{\rm c}$, $F_{\rm r}$ are governed by the same temperature gradient. Related to this, the Boussinesq model has one {parameter less}: the pressure scale height $H$. Because of this difference, the  heat flux in the stellar (compressible) model cannot be compared directly with  the heat flux in an incompressible model. Instead, the ratio of convective to radiative heat flux in the Boussinesq model  is to be identified with the ratio $ f=F_{\rm c}/F_{\rm rs}$ of the convective flux to the {superadiabatic component} of the radiative flux  in the star, in the limit $H\rightarrow\infty$ (cf.\ Massaguer \& Zahn 1980). The Boussinesq model thus is the limit  $\epsilon \downarrow 0$ (taken at fixed $\ras$). If the semiconvective layer thickness $\epsilon$ is `sufficiently small', the compressible case can be compared directly with the Boussinesq model {(as verified by the numerical tests with $\epsilon=1$ and $\epsilon=0.1$ in Sect.\   \ref{compar} above)}.

This makes the semiconvection problem more amenable to numerical simulation. In contrast with a convective stellar envelope for example, with its many scale heights to be covered, the layered nature of double diffusive convection puts it in a parameter range that is much closer to conditions accessible with realistic ab initio calculations.  

With (\ref{frad}) for the radiative flux, the total heat flux can be written as
\beq F=F_\mr{r}+F_\mr{conv}= k\nabla_{\rm a} + \mr{Nu}\, k(\nabla-\nabla_{\rm a}),\label{defnu}\eeq
where Nu is the Nusselt number. By taking out the radiative flux associated with the adiabatic gradient, the Nusselt number as defined in (\ref{defnu})  can now be identified with the Boussinesq equivalent (see also Massaguer \& Zahn 1980). The relevant Nusselt Number is thus not simply the ratio of total to radiative heat flux{\footnote{The implications of this distinction are not apparent in some of the published work on semiconvection.}.

\subsection{Rayleigh number and density ratio}

The modified Rayleigh number for a layer of thickness $d$ is, in astrophysical notation:
\beq \mr{Ra_*}=\mr{Pr\,Ra}={g d^4\over\kappa^2H}(\nabla-\nabla_\mr{a}),\label{raa}\eeq
where $\kappa$ is the thermal diffusivity (cm$^2$/s) and $g$ the acceleration of gravity. {\rast\  is typically a very large number unless the layer thickness $d$ is small (for estimates see the 15 $M_\odot$ example below).}
Apart from Le, the problem also depends on the relative strength of the stabilizing solute gradient relative to the destabilizing thermal gradient. This can be measured in terms of the thermal and solute buoyancy frequencies $N_{\rm T}$, $N_{\rm S}$:
\beq N_{\rm T}^2={g\over H}(\nabla_{\rm a}-\nabla),\eeq
\beq N_{\rm S}^2={g\over H}\,\nabla_\mu,\eeq
with
\beq \nabla_\mu\equiv{\rm d}\ln\mu/{\rm d} \ln P,\eeq
where $\mu$ is the mean atomic weight per particle,   and the gradients are understood as mean values over a double diffusive step.   The {density ratio} $R_\rho$:
\beq R_\rho\equiv-N_{\rm S}^2/N_{\rm T}^2 ={\nabla_\mu\over \nabla-\nabla_{\rm a} }\label{rrho}\eeq
is then the equivalent of the density ratio in the Boussinesq formulation, and is the dimensionless measure we will use for the relative strength of the stabilizing Helium gradient. Semiconvection,  i.e. a stratification `between Schwarzschild and Ledoux' then corresponds to $R_\rho>1$.      $ \nabla-\nabla_{\rm a} $ is typically a small number in a stellar interior (even in the presence of the double diffusive steps), and $R_\rho$ a large number (see below).

\subsection{Extrapolation}
\label{mixing}

The connection between the numerical results and the astrophysical conditions requires extrapolation  that cannot at present be covered by the simulations.  It is covered by the model in S13, as validated by the comparison with our numerical results. In contrast with the laboratory setup where {the temperature difference} is the imposed quantity,   the heat flux is the fixed quantity under the conditions in a stellar semiconvective zone, since the structure of the star depends only little on the behavior of the semiconvective zone. The asymptotic conditions for {large \rast} then lead (S13 Sect.\  6) to simple expressions for the superadiabaticity $\nabla-\nabla_\mr{a}$ and the effective Helium diffusivity $\kse$:
\beq \nabla-\nabla_\mr{a}\approx\mr{Le}^{1/2}\nabla_\mu,\label{sa}\eeq
\beq \kse=(\kappa_\mr{s}\kappa)^{1/2}{(\nabla_\mr{r}-\nabla_\mr{a})\over\nabla_\mu}\label{nusef}\eeq
(S13 eqs. 57, 59).  The analysis in S13 does not take into account a modification due to the effect of radiation pressure discussed in S92. With this modification, (\ref{nusef}) becomes:
\beq \kse=(\kappa_\mr{s}\kappa)^{1/2}({4\over\beta}-3){(\nabla_\mr{r}-\nabla_\mr{a})\over\nabla_\mu},\label{nusefc}\eeq
where $\beta=P_\mr{g}/(P_\mr{g}+P_\mr{r})$ is the ratio of gas to total pressure. The density ratio approaches the value
\beq R_\rho\approx \mr{Le}^{-1/2},\label{rra}\eeq
while the relative thickness of the stagnant zones in this limit is  of order (setting $\beta=1$)
\beq \delta\approx 1/\mr{Nu_s}=\mr{Le}^{1/2}\nabla_\mu/(\nabla_\mr{r}-\nabla_\mr{a}).\label{del}\eeq
The range of validity of the limit is
\beq l_0\ll d\ll H,\label{limit}\eeq
where
\beq l_0=(\kappa^2H/g)^{1/4} \label{l0}\eeq
is the length scale on which the thermal diffusion time scale equals the free fall time over a pressure scale height. 

{The asymptotic value $R_\rho= \mr{Le}^{-1/2}$ (eq.\ \ref{rra}) is also the maximum density ratio for which the theory predicts existence of the layered state. Semiconvection in a stellar model with heat flux fixed is thus predicted to settle close to this maximum, if the layer thickness is not small ($d\gg d_0$ or $\ras\gg 1$). How close it actually settles can also be predicted, but requires analysis to next order in the small quantity $1/\ras$.}

In a stellar evolution model, expressions (\ref{sa},\ref{nusefc}) give nearly identical results as the model of {Zaussinger (2011), which is based on S92}. They are easier to implement since they have been restricted to conditions $d\gg l_0$ that are the most interesting for stellar evolution anyway.

\subsection{A 15 $M_\odot$ star}
\label{stars}
We are now in the position to estimate  the range of parameter values for semiconvection in a star. Consider  the important case of massive stars around main sequence turnoff. We use a model  kindly provided by A.~Weiss (model `fzm15\_151'). Characteristic values for the physical quantities in the semiconvective zone {of} this model are $g\approx 10^6$ cm/s$^2$, $\ks \approx 1$ cm$^2$/s, $\kappa \approx 3\cdot10^8$ cm$^2$/s, $H\approx 2\cdot10^{10}$ cm, $\nabla_{\rm a}=0.4$, $\nabla_{\rm r}-\nabla_{\rm a}\approx 0.02$, $\nabla_\mu\approx 1$. 

{With (\ref{raa}) and (\ref{sa}), the Rayleigh number can be found for this semiconvective zone, as a function of the layer thickness $d$. This yields $\ras\sim 10^{12}$ for  $d/H=0.1$,  or $\ras\sim 10^{8}$ for $d/H=0.01$, for example. For such `macroscopic' layer thicknesses, conditions are thus in  the asymptotic regime for which the expressions in \ref{mixing} are approximately valid.}

The effective He-diffusivity from (\ref{nusefc}) is $ \kse\approx 10^3$ cm$^2$/s,   three orders of magnitude above the microscopic value.   The mixing time scale over a pressure scale height is thus about $10^{10}$ yr. The value of $l_0$ is $2\,10^5$ cm, or about $10^{-5}$ pressure scale heights. 
The limiting expressions  (\ref{sa}-\ref{del})  are thus applicable over a large range in the value of the (uncertain) layer thickness. The lower end of this range is not likely to {be} very relevant, since at $d\sim l_0$ the expected time scale for layer merging (see below) is very short, of the order of Le$^{-1/2}$ times the free fall time scale over a pressure scale height (from Eqs.\  \ref{l0}, \ref{thickn}).

{The time scale for the initial layer formation from a Kato oscillation is similarly short, a few times the instability growth time, which is of the order of the convective overturning time in the absence of the stabilizing $\mu$-gradient. For the 15 $M_\odot$ star this works out to about one day.}
{The asymptotic expressions (\ref{sa}) and (\ref{nusefc}) should therefore be adequate for use in stellar evolution codes.}

\subsection{Layer thickness}
\label{merge}
While the layer thickness has little {influence} on the effective mixing rate, it is useful to check if it could become significant compared with the pressure scale height,  in which case the limit of thin layering assumed would not be valid. This requires a model for the layer thickness,  for which no good theory is available. Observations in laboratory experiments and in geophysical cases show that layer thickness is not constant, but grows in time by processes of merging of neighboring layers {(McDougall 1981, Ross \& Lavery 2009, see also the numerical experiment in Young \& Rosner 2000)}. Layer thickness can therefore \textit{not be treated independently of the history of the system}}. 

An estimate can be made, however, from the effective solute diffusivity. Changes in layer thickness involve  the exchange of solute between neighboring layers. Over a distance $D$, the time scale $\tau$ for exchange is given by the effective diffusivity $ \kse$: $\tau\approx D^2/ \kse$. Setting $D$ equal to the layer thickness $d$ itself then gives an estimate of the rate of change of thickness by merging:
\beq \mr{d}\ln d/\mr{d}t=1/\tau,\eeq
or
\beq {\mr{d}\over\mr{d}t} d= {\kse\over d}.\eeq
If $ \kse$ is constant in time:
\beq d=(2 \kse t)^{1/2},\label{thickn}\eeq
where $t$ is time since the semiconvective condition started. In the $15\, M_\odot$ star the duration of the semiconvective state is of {the} order {of} $10^7$ yr, so the layer thickness to be expected at the end of the semiconvective phase is of the order $10^9$ cm, or $0.05\, H$. Given the uncertainties involved, it cannot be ruled out that semiconvective layer thickness $d$ can actually become {`macroscopic', of} order $H$, in the course of the star's evolution.

\section{Discussion}
\label{discussion}

As in the case of ordinary convection, there are far fewer intrinsic parameters in semiconvection than the quantities defining the physical state in a stellar interior. This allows a significant volume of astrophysically useful parameter space to be covered by a grid of numerical simulations. A physical model as applied here is still needed, however, to interpret the results and to extrapolate them meaningfully to the  astrophysically relevant range.  

The low value of the viscosity and solute diffusivity compared with the thermal diffusivity constitute a limiting case that actually simplifies the double diffusive problem greatly. Among other things, the results become nearly independent of the Prandtl number in this limit {[as suggested already by Proctor's (1981) analysis]}. 

The simulations were all done in 2-D, so tests in 3-D will be needed for verification. It is  unlikely, however, that the results will turn out very different,  at least within an astrophysical factor of 2. The reason is that in the low-Pr, low-Le limit the flow in the layers is almost equivalent to ordinary convection between plates. Due to the low solute diffusivity, the amount of solute in the bulk of the layer is small. It can then be treated as a perturbation, as assumed  in S92. Known laboratory results for convection at very high Rayleigh numbers can then be used to extrapolate the numerical results. As shown in S13, this makes predictions similar to the simple 2-D model used in S92. In particular the effective mixing rate is very low,  as also found (in a different region in parameter space) in geophysical examples like lake Kivu {and the (ant)arctic oceans} (cf. Sect.\  \ref{semidd} and S13).  

Stochastic fluctuations in the flow produce scatter in the fluxes of heat and solute, which affect the measured averages (cf.\ Table 1). The accuracy of these averages could be improved with longer runs.

Most of the results presented are based on simulation of a single double diffusive layer. As we found in \ref{double}, this does not fully reproduce broadening of the interfaces between layers by surface waves (see movie at Fig.\ \ref{fig:ds1}). 

The physics determining the thickness of the individual double diffusive layers remains uncertain. In the equivalent geophysical examples semiconvective zones always consist of a large number of  very long-lived layers.  Observations in the east-african rift lakes (e.g. Schmid et al. 2010) show that layers first forming at the boundary of an expanding double-diffusive zone are always thin, subsequently growing slowly by a process of merging.  This is understood in terms of an energy argument (cf. Sect.\  \ref{semidd}),  and is likely to happen in a growing stellar semiconvective zone as well. 

The estimate (eq.  \ref{thickn}) suggests that layer thickness might approach a significant fraction of a pressure scale height.  In this case, only a few layers would be left at the end of the semiconvective phase, and the location of their boundaries may well vary somewhat randomly between stars. As surmised in Zaussinger (2011), this might introduce a random element in the late stages of the evolution of massive stars.

{Semiconvection as studied here addresses only one of two astrophysical meanings of the term. In addition to the effect of a stabilizing solute gradient, there is the effect of a composition-dependent opacity: the increase of the radiative gradient when mixing is assumed, with consequent onset of convection in a stratification which was radiative before mixing. Historically, this has actually been the main concern in stellar evolution computations. The physics of this kind of semiconvection is completely different from the double-diffusive kind (cf.\ Kippenhahn et al. 2013); it has not received the same theoretical attention.}

\section{Conclusions}

The results show how the {physics of double diffusive convection known from geophysical and laboratory studies} can be applied to the astrophysical case of low Prandtl and Lewis numbers. As expected, the process takes place in the form of the  characteristic double diffusive layering known and theoretically understood since the 1980's. 

In the asymptotic regime occupied by astrophysical conditions the number of independent parameters determining the physics is effectively reduced to three, so that a meaningful range of parameter values can be covered with numerical simulations. We have compared a grid of simulations with the predictions of the model for a semiconvective layer in S13. In the parameter region where the estimate overlaps with the region covered by the simulations we find good agreement, even without significant tuning of the one fitting parameter used (the `erosion factor' $q$, Sect. \ref{sol}).

 The simulations also provide an answer to an interesting question raised by Proctor's (1981) mathematical analysis and by the model in S13. Both predicted that a critical number $R_{\rho}={\rm Le}^{-1/2}$ plays a role in the steady layered state. The results presented in Sect. \ref{max} clarify the behavior of the layered state around this maximum density ratio. In a system set up in a layered state just above $R_{\rho\rm\, max}$ the Nusselt numbers show large fluctuations in amplitude; the duration of the fluctuations becomes increasingly long as  $R_\rho$ approaches $R_{\rho\rm\, max}$. This explains why theories based on the assumption of stationarity fail near this limit.

Boussinesq and fully compressible simulations give equivalent results in the limit of small layer thickness. The Boussinesq calculations even reproduce the approximate Nusselt numbers for layers as thick as a scale height. The numerical results confirm the theoretical expectation that at low Prandtl number the dependence on viscosity is weak, and extends to Pr of {order} unity. 

The effective mixing rate in the semiconvective zone of a $15\, M_\odot$ main sequence star is predicted to be low,  though the semiconvective phase may possibly leave significant jumps in the profile of Helium concentration.

\begin{acknowledgements} 
We thank the referee for his perceptive and detailed review, which has led to substantial improvements.  F.\ Zaussinger is grateful to F.\ Kupka and H.J.\ Muthsam for discussions 
on the numerical solution of binary mixture equations with high resolution methods. He was 
supported by the DFG within the project `Modelling of
diffusive and double-diffusive convection' (projects KU 1954-3/1 and KU 1954-3/2 in SPP 1276/1 and SPP 1276/2, project leader F.\ Kupka) within the interdisciplinary Metstroem project. He was also supported by the Austrian Science 
Foundation (project P20973, Numerical Modelling of Semiconvection, project leader 
H.J.\ Muthsam). We also thank H.\ Grimm-Strele for the implementation of a parallel Poisson solver in the ANTARES code suite.
\end{acknowledgements}

%\bibliographystyle{plain}
%\bibliography{fzbib}

\begin{thebibliography}{99}
%\input{refs}

\bibitem{} Biello, J.~A.\ 2001, Ph.D.~Thesis,  U. of Chicago

\bibitem{}Castaing, B., Gunaratne, G., Kadanoff, L., Libchaber, A., 
\& Heslot, F.\ 1989, Journal of Fluid Mechanics, 204, 1 

\bibitem{}Huppert, H.E., \& Turner, J.S., 1981, Journal of Fluid Mechanics, 106, 299

\bibitem{}Kato, S.\ 1966, PASJ 18, 374

\bibitem{}Kippenhahn, R.,  Weigert, A.\  \& Weiss, A., 2013, Stellar Evolution, Springer Berlin (ISBN 978-3-642-30255-8)

\bibitem{}Langer, N., El Eid, M.~F., \& Fricke, K.~J.\ 1985, \aap, 145, 179 

\bibitem{}Langer, N., El Eid, M.~F., \& Baraffe, I.\ 1989, \aap, 224, L17 

\bibitem{}Linden, P.~F., \& Shirtcliffe, T.~G.~L.\ 1978, Journal of Fluid Mechanics, 87, 417 

\bibitem{}Lu, H., Swift, H.P.\ 2001, in Conference on desalination strategies, Jerba 11-12 Sept 2000, Elsevier (available at www.desline.com/articoli/4052.pdf)

\bibitem{} Maeder, A.\ 1997, \aap, 321, 134 

\bibitem{}Massaguer, J.M. \& Zahn, J.-P.\ 1980 A\&A 87, 315

\bibitem{}McDougall, T.\ 1981, Progress in Oceanography, 10, 91 

\bibitem{}Merryfield, W.~J.\ 1995, \apj, 444, 318 

\bibitem{}Muthsam, H.~J., Kupka, F., L{\"o}w-Baselli, B., Obertscheider, C., Langer, M., \& Lenz, P.\ 2010, New Astronomy 15, 460 

\bibitem{}Muthsam, H.~J., Goeb, W., Kupka, F., Liebich, W., \& Zoechling, J.\ 1995, A\&A 293, 127

\bibitem{}Muthsam, H.~J., Goeb, W., Kupka, F., \& Liebich, W. \ 1999, New Astronomy 4, 405

\bibitem{}Nayar, A.\ 2009, Nature 460, 321

\bibitem{}Niemela, J.J., Skrbek, L., Sreenivasan, K.R., \& Donnelly R.J.\  2000, Nature 404, 837 

\bibitem{}Padman, L. \& Dillon, T.M., 1987, \jgr, 92, 10799

\bibitem{}Proctor, M.R.E.\ 1981, Journal of Fluid Mechanics, 105, 507

\bibitem{}Rosenblum, E., Garaud, P., Traxler, A., \& Stellmach, S.\ 2011, \apj, 731, 66 

\bibitem{}Ross, T., \& Lavery, A.\ 2009, Experiments in Fluids, 46, 355 

\bibitem{}Schladow, S.~G., \& Imberger, J.\ 1987, \jgr, 92, 6501 

\bibitem{}Schmid, M., Busbridge, M., W\"ust, A.\ 2010, Limnol.\ Oceanogr.\ 55, 225

\bibitem{}Schmitt, R.\ 1987, Deep Sea Research Part I: Oceanographic Research, 34, 1655 

\bibitem{}Schmitt, R.~W.\ 1994, Annual Review of Fluid Mechanics, 26, 255 

\bibitem{}Shirtcliffe, T.\ G.\ L.\, 1967, Nature 213, 489

\bibitem{}Shu, C.-W., \& Osher, S.\ 1988, Journal of Computational Physics 77, 439 

\bibitem{}Sigvaldason, G.\ 1989, Journal of Volcanology and Geothermal Research, 39, 97 

\bibitem{}Spiegel, E.~A.\ 1969, Comments on Astrophysics and Space Physics, 1, 57 

\bibitem{}Spiegel, E.~A.\ 1972, \araa,, 10, 269 

\bibitem{}Spruit, H.C.\ 1992, A\&A 253, 131 (S92)

\bibitem{}Spruit, H.C.\ 2013, accepted A\&A (S13)

\bibitem{}Tayler, R.J.\ 1953,  Doctoral Dissertation, Cambridge University

\bibitem{}Turner, J.~S., \& Stommel, H.\ 1964, Proceedings of the National Academy of Science, 52, 49 

\bibitem{}Turner, J.S.\ 1979, Buoyancy Effects in Fluids, Cambridge University Press

\bibitem{}Turner, J.S.\ 1985, Ann. Rev. Fluid Mech. 17, 11

\bibitem{} Stevenson, D.~J.\ 1979, \mnras, 187, 129

\bibitem{} Veronis, G.\ 1968, Journal of Fluid Mechanics, 34, 315 

\bibitem{}Young, Y., \& Rosner, R.\ 2000, \pre, 61, 2676 

\bibitem{} Zaussinger, F.\ 2011, PhD thesis, University of Vienna, Numerical simulation of double-diffusive convection, http://pubman.mpdl.mpg.de/pubman/item/escidoc:1121652:2 

\end{thebibliography}

\end{document}